\documentclass[twocolumn,floatfix,pra,showpacs,superscriptaddress]{revtex4}

\usepackage{color}
\usepackage{array}
\usepackage{amsmath}
\usepackage{amssymb}
\usepackage{wasysym}
\usepackage{bm}
\usepackage{graphicx}
\usepackage{txfonts}
\usepackage{epsfig}

\newcommand{\abs}[1]{|#1|}
\newcommand{\bra}[1]{\langle \, #1 \,|}
\newcommand{\ket}[1]{|\, #1 \, \rangle}
\newcommand{\bket}[2]{\langle \, #1 \,|\, #2 \, \rangle}
\newcommand{\boket}[3]{\langle\, #1 \,|\, #2 \,|\, #3 \,\rangle}
\newcommand{\ketbra}[1]{\ket{#1} \bra{#1}}

\newcommand{\be}{\begin{equation}}
\newcommand{\ee}{\end{equation}}
\newcommand{\bc}{\begin{center}}
\newcommand{\ec}{\end{center}}

\newcommand{\non}{\nonumber}

\begin{document}
\title{Measurement of many-body chaos using a quantum clock}

\author{Guanyu Zhu}
\affiliation{Joint Quantum Institute, NIST/University of Maryland, College Park, MD 20742, USA}

\author{Mohammad Hafezi}
\affiliation{Joint Quantum Institute, NIST/University of Maryland, College Park, MD 20742, USA}
\affiliation{Kavli Institute of Theoretical Physics, Santa Barbara, CA 93106, USA}
\affiliation{Department of Electrical and Computer Engineering and Institute for Research in Electronics and Applied Physics,
University of Maryland, College Park, MD 20742, USA}

\author{Tarun Grover}
\affiliation{Department of Physics, University of California at San Diego, La Jolla, CA 92093, USA}
\affiliation{Kavli Institute of Theoretical Physics, Santa Barbara, CA 93106, USA}

\begin{abstract}

There has been recent progress in understanding chaotic features in many-body quantum systems. Motivated by the scrambling of information in black holes, it has been suggested that  the time dependence of out-of-time-ordered (OTO) correlation functions such as $\langle O_2(t) O_1(0) O_2(t) O_1(0) \rangle $ is a faithful measure of quantum chaos. Experimentally, these correlators are challenging to access since they apparently require access to both forward and backward time evolution with the system Hamiltonian. Here, we propose a protocol to measure such OTO correlators using an ancilla which controls the direction of time. Specifically, by coupling the state of ancilla to the system Hamiltonian of interest, we can emulate the forward and backward time propagation, where the ancilla plays the role of a `quantum clock'.  Within this scheme, the continuous evolution of the entire system  (the system of interest and the ancilla)  is governed by a time-independent Hamiltonian. Our protocol is immune to errors that could occur when the direction of time evolution is externally controlled by a classical switch.
\end{abstract}
\pacs{03.67.-a, 42.50.Pq, 05.45.Mt, 05.30.-d}
\maketitle

\section{Introduction}
\label{sec:intro}

Characterizing chaos in single-particle quantum systems is an old and rich topic with roots in semiclassical quantization \cite{Gutzwiller:1990}. However, there is relatively less understanding of chaos in many-body quantum systems and quantum field theories, especially away from the semi-classical limit.
Recently, new progress has been obtained in characterizing chaos in quantum many-body systems using insights from the scrambling of information in black holes \cite{Shenker:2014, Kitaev:2014, Maldacena:2015}. Specifically, it has been argued that the time dependence of the four-point correlation function $\langle O_2(t) O_1(0) O_2(t) O_1(0) \rangle$ involving any two local operators $O_1, O_2$  is a  measure of quantum chaos, where the averaging $\langle \rangle$ denotes averaging over a canonical ensemble \cite{footnote1}. 
Since this correlator is the overlap between two states that are obtained by applying the non-commuting operators $O_1(0)$ and $O_2(0)$ in reverse orders with respect to each other, the basic intuition  is that it captures the sensitivity of the evolved system to initial conditions \cite{footnote2}. 
Remarkably, it has been shown by Maldacena, Shenker, and Stanford \cite{Maldacena:2015} that such a correlator can not grow faster than $e^{\lambda t}$, with a universal bound $\lambda \leq 2\pi T/\hbar$, thus defining a maximal `Lyapunov exponent'. Even more, the bound is known to be saturated by certain large-$N$ conformal field theories which are holographically described by the Einstein's gravity \cite{Roberts:2015, Shenker:2014,Roberts:2015_2}, and also \cite{Polchinski:2016, Maldacena:2016} by a non-local Hamiltonian (`SYK model'), originally discussed by Sachdev and Ye \cite{Sachdev:1994}, and more recently in the context of holography by Sachdev \cite{Sachdev:2010}, and Kitaev \cite{Kitaev:2014}.

The peculiar feature of the correlator $\langle O_2(t)O_1(0) O_2(t) O_1(0) \rangle$ is that it is not time-ordered and thus, from an experimental point of view, requires access to time evolution by a Hamiltonian $H$, \textit{and} $-H$. Recently, an interesting proposal was made in Ref.\cite{Swingle:2016td} where they outlined a protocol to measure $\langle O_2(t) O_1(0) O_2(t) O_1(0) \rangle$. The sign of the Hamiltonian in Ref.\cite{Swingle:2016td} is changed via a `classical switch' -- by noticing that the sign of interactions in a cavity QED depends on the sign of the two-photon detuning, it was proposed that it can be changed \textit{in-situ}.  However, any imperfection in this sign reversal due to experimental imperfections could lead to significants error in OTO correlators.

In this paper, we address this issue by proposing a `quantum clock' to  control the sign of a certain many-body Hamiltonian and use it to construct a new proposal to measure OTO correlators such as $\langle O_2(t)O_1(0) O_2(t) O_1(0) \rangle$. The basic idea is to couple the Hamiltonian $H$ of interest to an ancilla qubit $\vec{\tau}$ linearly such that $H_\text{tot} = \tau^z \otimes  H $, and then perform a unitary time evolution of the state $(|\uparrow\rangle + |\downarrow\rangle) \otimes |\psi\rangle_S  $ where $|\psi\rangle_S$ is some initial state of the system of interest. By construction of $H_\text{tot}$, the $|\uparrow\rangle$ branch of the wavefunction effectively evolves forward in time while the $|\downarrow\rangle$ branch evolves backward. Therefore, the ancilla qubit $\vec{\tau}$ effectively acts like a `quantum clock'  which controls the direction of time evolution.  
The OTO correlator $\langle O_2(t) O_1(0) O_2(t) O_1(0) \rangle$ is then measured by conditionally applying different operators on the forward and backward evolving branches of the wavefunction, and finally measuring the expectation value of the operator $\tau^x$ acting on the quantum clock. 

From an experimental standpoint, our protocol is motivated by the rapid development of quantum simulation and information technology in recent years, such as cavity quantum electrodynamics (QED) \cite{birnbaum_photon_2005, Jiang:2008gs, Douglas:2015hda, lezTudela:2015gd}, circuit-QED \cite{Schoelkopf:2008vi, Blais:2007hh, houck2012, koch_time-reversal_2010, Hoffman:2011fz, underwood2012, Schmidt:2013us,  Raftery:2014jk, Chiesa:2015vo, HacohenGourgy:2015th}, Rydberg atoms \cite{Lukin:2003ct, Saffman:2010ky, Anonymous:s6xSQwvw, Sommer:2015ur}, and trapped ions \cite{Kim:2010ib}; it is within current technology to engineer an ancilla qubit coupled to a many-body system globally. The ancilla qubit can be either the cavity photon mode  or the internal state of an  atom.  The mechanism of the coupling is usually through dispersive interaction, which can originate, for example from Jaynes-Cummings interaction \cite{Jaynes:1963fa} perturbatively \cite{Blais:2007hh, Jiang:2008gs, dicarlo_demonstration_2009}, or from Rydberg blockade mechanism \cite{Saffman:2010ky, Comparat:2010cb, Hofmann:2013gm, Maller:2015is}.  Such an ancilla has been widely used as control-phase gate \cite{dicarlo_demonstration_2009, Saffman:2010ky, Comparat:2010cb, Hofmann:2013gm, Maller:2015is} for quantum information processing.  Meanwhile, theoretical proposals suggest that such an ancilla can be used as a quantum switch that performs a many-body Ramsey interferometer \cite{Muller:2009fp,Knap:2013jy} to extract useful information of the quantum system, such as entanglement entropy \cite{Abanin:2012wd} and spectrum \cite{Pichler:2016ua}.   In this paper, the ancilla, in addition to playing the role of the quantum clock, has the added benefit of being the probe of the system. Specifically, we show how the OTO could be obtained by measuring the same ancilla. 

The primary advantage of our protocol utilizing a quantum clock for both control and readout of the many-body states is its robustness against statistical errors, such as imperfect rotation, in each shot of the experiments.  In particular, our quantum clock does not modify the many-body Hamiltonian \textit{in-situ}, which is in contrast to a previous proposal of measuring the same correlator using a classical switch to continuously tune the prefactor of the Hamiltonian \cite{Swingle:2016td}.   In addition,  we are also able to construct a local Hamiltonian, which is more physical from condensed matter and quantum field theoretic viewpoint, and may also exhibit richer behavior of quantum scrambling.

We also provide simple examples of embedding such an ancilla in cavity-QED systems, for both a non-local all-to-all coupled spin models and a local XY-spin or extended Bose-Hubbard model.  In the non-local model, qubits (spins) are interacting with each other mediated by a passive cavity bus, which is itself dispersively coupled to another ancilla cavity in order to control the sign of the Hamiltonian.  To realize the local model, local cavities/resonators are coupled by intermediate qubits, which are themselves coupled to a global cavity.  When integrating out the qubit degrees of freedom and with proper choosing of parameters,  the effective Hamiltonian has an overall sign controlled by the state of the global cavity.
Such models can be realized with recently developed experimental platforms such as circuit-QED network \cite{houck2012, koch_time-reversal_2010, Hoffman:2011fz, underwood2012, Zhu:2013cm, Raftery:2014jk, Chiesa:2015vo} and qubit/atomic array in a 3D cavity \cite{HacohenGourgy:2015th}.

The outline of our paper is as follows.  In Sec.~\ref{general},  we present our general protocol of measuring the OTO correlator with a quantum clock.  In Sec.~\ref{models}, we show how such a quantum clock could be embedded in a physical model.  In Sec.~\ref{circuit-QEDprotocol}, we discuss the implementation of the protocol with circuit-QED systems.  In Sec.~\ref{error}, we analyze the stability of our protocol against imperfections.  We present the generalization of approach for extended Bose-Hubbard model and disordered spin chains in Sec.~\ref{sec:generalization}. We provide the conclusion and outlook in Sec.~\ref{conc}.   In Appendix~\ref{sequence}, we list the complete sequence of gates in the protocol.  We show the details of the experimental realization of the local model which we construct in the main text with a circuit-QED network or a qubit array in a 3D cavity in Appendix~\ref{circuit-QED}.  In Appendix \ref{sec:comparison}, we compare the numerical diagonalization of the original and second-order effective Hamiltonian. Finally, in Appendix~\ref{complete}, we provide a complete formula of the second-order dispersive Hamiltonian we mention in Sec.~\ref{models} without integrating out the qubits.  

\section{General scheme}\label{general}
We consider a many-body system governed by Hamiltonian $H$ and couple it globally to an ancilla qubit $\tau^z$, with the total Hamiltonian being
\be\label{ancilla}
H_\text{tot} = \tau^z \otimes H.
\ee 
With the cavity-QED implementation, the ancilla qubit can also be realized with the global cavity photon mode as $\tau^z$$=$$1$$-$$2 a^\dag a$, if the cavity photon state is restricted in the $0$- and $1$-photon subspace. Hence the total Hamiltonian of the coupled system can also be expressed as
\be\label{Htot}
H_\text{tot} = (1-2a^\dag a ) \otimes H.
\ee
From now on, we call both the cavity and the ancilla qubit as `ancilla' without further specification, since they play the same role and one can use either them for the protocol.

In Eq.~\eqref{ancilla} and \eqref{Htot}, the ancilla only dresses the many-body system $H$, and does not exchange excitations (photons) with the many-body system. Crucially, if the $H$ we consider is a local Hamiltonian, the ancilla does not mediate long-range interaction between the particles/spins in the many-body system and preserves the locality of $H$.  

The only thing that the ancilla does is to control the overall sign of the many-body Hamiltonian $H$ quantum coherently.  If the cavity contains no photon, namely the ancilla is in state $\ket{0_a}$ \cite{footnote3}, the overall sign is `$+$'; if the cavity contains one photon, namely the ancilla is in state $\ket{1_a}$, the overall sign is `$-$'.  If we consider the dynamics of the coupled system, we can express the evolution operator as 
\be\label{evolution}
U_\text{tot}(t)=e^{-i H_\text{tot} t} = e^{-i H t} \otimes \ketbra{0_a} + e^{i H t} \otimes \ketbra{1_a}.  
\ee
This means that the many-body system $H$ evolves forward in time if the cavity contains no photon, and backward in time if the cavity contains one photon.  Namely the cavity photon number $a^\dag a$ or the ancilla qubit $\tau^z$ acts a binary `quantum clock' that controls  the `arrow of time'.  More interestingly since the `clock' is quantum, the system can be in a parallel superposition of evolving both forward and backward in time, for example when we prepare the `clock' being in the superposition state $\frac{1}{\sqrt{2}}(\ket{0_a}+\ket{1_a})$.

Now we discuss a general protocol to measure the out-of-time-order (OTO) correlator $\langle O_2(t) O_1(0)  O_2(t) O_1(0) \rangle$ introduced earlier, where $O_1$ and $O_2$ are certain operators, and $O(t) = e^{i H t} O e^{-i Ht}$ is the Heisenberg evolved operator.  The average `$\langle \rangle$' could be with respect to a certain initial state $\ket{\psi}_S$ or an ensemble average over a thermal density matrix $\rho_S$$=$$\sum_S \frac{e^{-\beta H}}{Z} \ket{\psi}_S{}\,_S\bra{\psi}$, where $Z$ is the partition function. For the sake of convenience, we will focus on average with respect to a given pure state $|\psi\rangle_S$. If one is interested in average with respect to a thermal ensemble, one can still work with a pure state that is obtained by time-evolving an initial finite-energy density pure state with respect to $H$ \cite{Srednicki:1998}. Assuming that the system is generic (non-integrable), the pure state average is then expected to match the thermal ensemble average at a temperature determined by the energy density of the state \cite{Deutsch:1991, Srednicki:1994, Srednicki:1998}.

\begin{figure*}
\includegraphics[width=1.5\columnwidth]{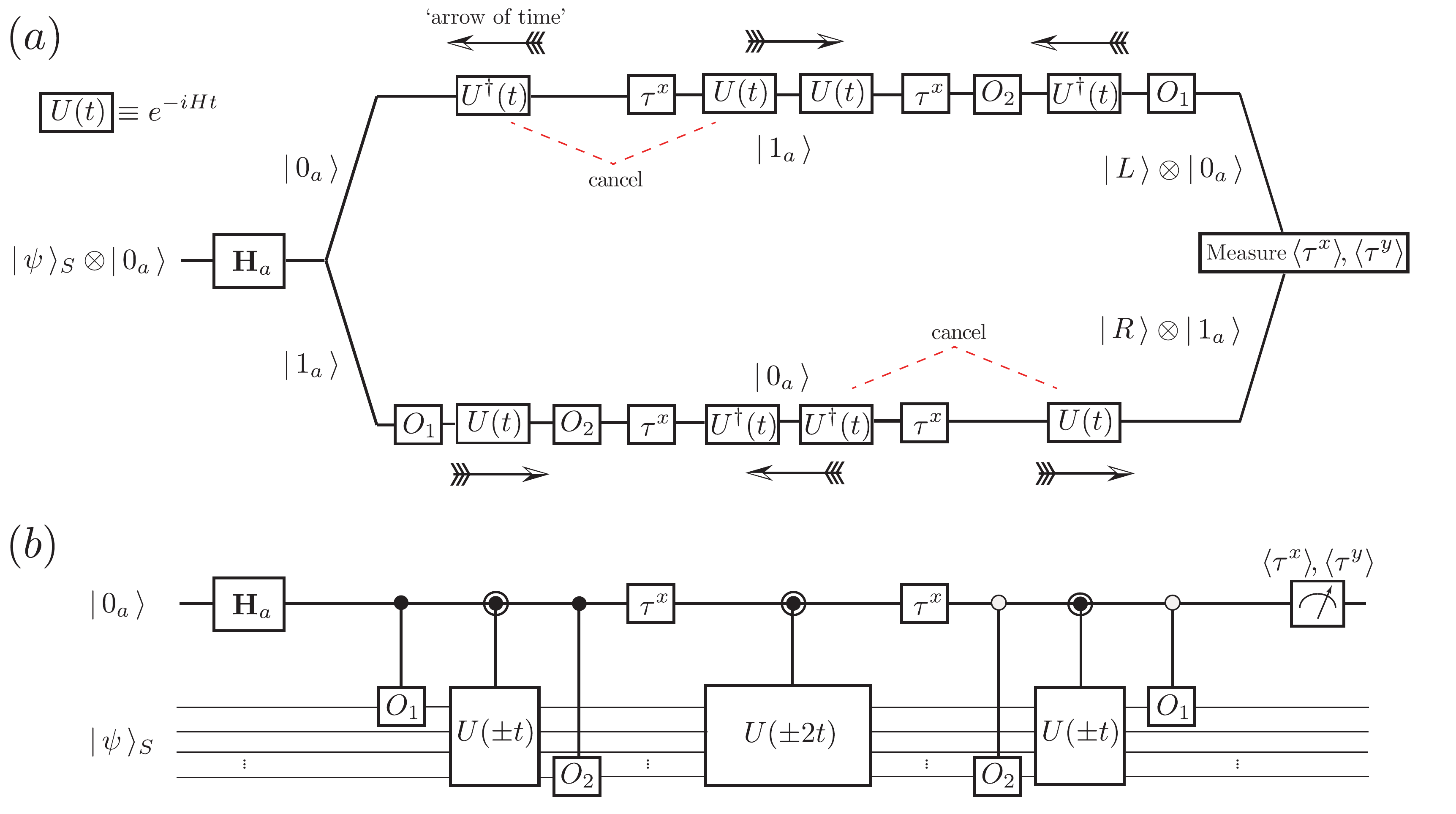}
\caption{(a) Illustration of the Ramsey interferometry protocol. The interferometry starts from the left, with the initial state $\ket{\psi}_S \otimes \ket{0_a}$.  The Hadamard rotation splits the time evolution of the many-body state $\ket{\psi}_S$ into two branches, conditioned by the ancilla. The time evolution conditioned by ancilla state $\ket{0_a}$ ($\ket{1_a}$) is forward (backward) in the beginning.   After applying the $\tau^x$ operations, the ancilla states on the two branches interchange, and so are the directions of time evolution.  The red  dashed lines show the canceled time evolution. Conditional operations $O_1$ and $O_2$ on either branch are applied. A final measurement of the ancilla in the x- and y-basis gives the real and imaginary part of the OTO correlator. We emphasize that the actual experimental time always goes from left to right. (b) The quantum circuit description of the same protocol. }
\label{circuitdaigram}
\end{figure*}

In the Schrodinger picture, the correlator corresponding to a particular initial state can be written as 
\be
\nonumber _S\bra{\psi}e^{i Ht} O_2 e^{-i Ht}  O_1 e^{i Ht}   O_2 e^{-i Ht} O_1 \ket{\psi}_S.
\ee
To measure this correlator, we apply the following Ramsey interferometry protocol as illustrated in Fig.~\ref{circuitdaigram}: 
\begin{enumerate}
\item
Start with the many-body system in the state $\ket{\psi}_S$ with respect to which we wish to measure the OTO correlator. Thus, the coupled system can expressed as $\ket{\psi}_S \otimes \ket{0_a}$.  
\item
Apply a Hadamard gate, i.e.~a $\pi/2$-rotation (pulse) around the $y$-axis to the ancilla state:


The coupled system is thus prepared in the superposed state
$ \nonumber \frac{1}{\sqrt{2}} \ket{\psi}_S \otimes [\ket{0_a}+\ket{1_a}].$
From now on, the evolution of the many-body system split into two branches, conditioned by the ancilla state $\ket{0_a}$ and $\ket{1_a}$ respectively.  
\item
Apply a conditional operation 
\be
C_{O_1, 1}= O_1 \otimes \ketbra{1_a} +\mathbb{I}_S \otimes \ketbra{0_a},
\ee
so that $O_1$ is  applied only to the lower branch of the interferometer conditioned by the ancilla state $\ket{1_a}$.
The coupled system forms an entangled state
\[
\frac{1}{\sqrt{2}} [ O_1 \ket{\psi}_S \otimes \ket{1_a}+ \ket{\psi}_S \otimes \ket{0_a}].
\]
\item
Let the system evolve with total Hamiltonian $H_\text{tot}$ for time $t$ according to $U_\text{tot}(t)$ represented in Eq.~\eqref{evolution}.   The coupled system is now in an entangled state of evolving forward and backward in time conditioned by the photon number, namely  
\be
\frac{1}{\sqrt{2}} [ e^{-i Ht} O_1 \ket{\psi}_S \otimes \ket{1_a}+ e^{i Ht} \ket{\psi}_S \otimes \ket{0_a}].
\ee
\item
Apply a conditional-$O_2$ on the lower ($\ket{1_a}$) branch:
\be
C_{O_2, 1}= O_2 \otimes \ketbra{1_a} +\mathbb{I}_S \otimes \ketbra{0_a}.
\ee

\item
In order to reverse the `arrow of time' in both branches, we simply apply a $\tau^x$ operator ($\pi$-pulse around the $x$-axis)  to flip the ancilla. Then we let the coupled system evolve for a period of $2t$ and reach the state
\be
\nonumber \frac{1}{\sqrt{2}} [e^{2 iHt} O_2 e^{-i Ht} O_1 \ket{\psi}_S \otimes \ket{0_a}+e^{-2 i Ht}  e^{i Ht} \ket{\psi}_S \otimes \ket{1_a}].
\ee

\item
Perform the previous steps (3-6) with reversal order (with conditioned operations on the other branch) as shown in Fig.~\ref{circuitdaigram}(a), the coupled system ends up with the final state
\begin{align}\label{final}
\ket{\Psi_f}=\frac{1}{\sqrt{2}} [\ket{R} \otimes \ket{1_a}+ \ket{L} \otimes \ket{0_a}],
\end{align}
where we have abbreviated the wavefunctions in two branches as
\be
\nonumber \ket{R} \equiv e^{i Ht} O_2 e^{-i Ht} O_1 \ket{\psi}_S,  \ \ket{L} \equiv O_1 e^{i Ht} O_2 e^{-i Ht} \ket{\psi}_S.
\ee

\item Measure the expectation value of $\tau^x$ operator under the final state  $\ket{\Psi_f}$, which effectively takes an overlap between the many-body states in the two branches of the interferometer and leads to
\begin{align}
\nonumber  &\langle \tau^x \rangle_f \equiv \boket{\Psi_f}{\mathbb{I}_S\otimes\tau^x}{\Psi_f}  = \text{Re}[ \bket{L}{R}] \\
=& \text{Re}[_S\bra{\psi}e^{i Ht} O_2 e^{-i Ht}  O_1 e^{i Ht}   O_2 e^{-i Ht} O_1 \ket{\psi}_S].
\end{align}
The outcome is the real part of the OTO correlator.  Similarly, one can extract the imaginary part by measuring $\tau^y$,  since $\langle \tau^y \rangle_f = \text{Im}[ \bket{L}{R}] $.  

\end{enumerate}
Note that a part of forward time evolution has been canceled with backward time evolution in both branches [as illustrated in Fig.~\ref{circuitdaigram}(a) by red dashed lines]. For a complete sequence of operations in the protocol, see App.~\ref{sequence}. The preparation of states $\ket{R}$ and $\ket{L}$ can be interpreted as two gedankenexperiments:  (I) apply $O_1$, wait for time $t$, apply $O_2$, go backward in time for $-t$;  (II) apply $O_2$ at time $t$,   go backward in time for $-t$, and apply $O_1$ (at an earlier time than applying $O_2$).   The OTO correlator takes the overlap between these two states and hence compare the sensitivity of the state to the order of applying $O_1$ and $O_2 (t)$, or equivalently the sensitivity to the initial condition, and hence characterizes the butterfly effect.


\section{Physical models}\label{models}

In this section, we first discuss the realization of a simple non-local model with all-to-all spin couplings, where the overall sign is controlled by a quantum clock. Next, we discuss a local lattice model, with nearest neighbor couplings. The former is easier to implement while the latter is more relevant in a condensed matter context. The advantage of non-local model is that it does not suffer from errors due to imperfection in couplings (see Sec.\ref{sec:error_quantum}), and furthermore, from a physics standpoint, all maximally chaotic models known so far are non-local \cite{Sachdev:1994, Sachdev:2010, Kitaev:2014}, and thus worth exploring.

\subsection{Non-local model}\label{sec:nonlocal}

The model consists of $N$ qubits (spins) located in a coupler cavity bus with Jaynes-Cummings (JC) interactions \cite{Jaynes:1963fa}. In addition, an ancilla/control cavity (quantum clock) is coupled to the coupler cavity dispersively.  The entire system Hamiltonian is $H_s =H_0 + V$ where
\begin{align}
\non H_0=& \omega_a a^\dag a + \omega_b b^\dag b + \sum_{j=1}^N \frac{1}{2}\epsilon  \sigma^z_j + \eta  a^\dag a  b^\dag b \\
V=&  \sum_j g_j (\sigma^+_j b + \text{H.c.})  \label{eq:Hnonlocal}
\end{align}
where $a^\dagger (b^\dagger) $ is the creation operator associated to the ancilla (coupling bus) and $\omega_a(\omega_b)$ are the corresponding frequencies. $\sigma_j^z $ is the $j$-th qubit operator and $\epsilon$ is the corresponding frequency.  We require these three frequencies to be detuned away from each other so that effectively, there is no exchange between different types of excitations.  In particular, we choose $\epsilon < \omega_b <\omega_a$.  We define the detuning between qubits ($\sigma_j$) and coupling cavity bus ($b$) as $\Delta_b = \epsilon - \omega_b$.  The last term in $H_0$ is the cross-Kerr interaction (with strength $\eta$) between the coupling ($b$) and ancilla ($a$) cavities, which can be experimentally realized, e.g., by coupling two superconducting cavities with a Josephson junction \cite{Jin:2013tz, Nigg:2012wh}.  In the JC interaction term $V$, $g_j$ is the interaction strength between cavity and system qubits, which in general can depend on the qubits' locations and can also be disordered.

\begin{figure}
\includegraphics[width=1\columnwidth]{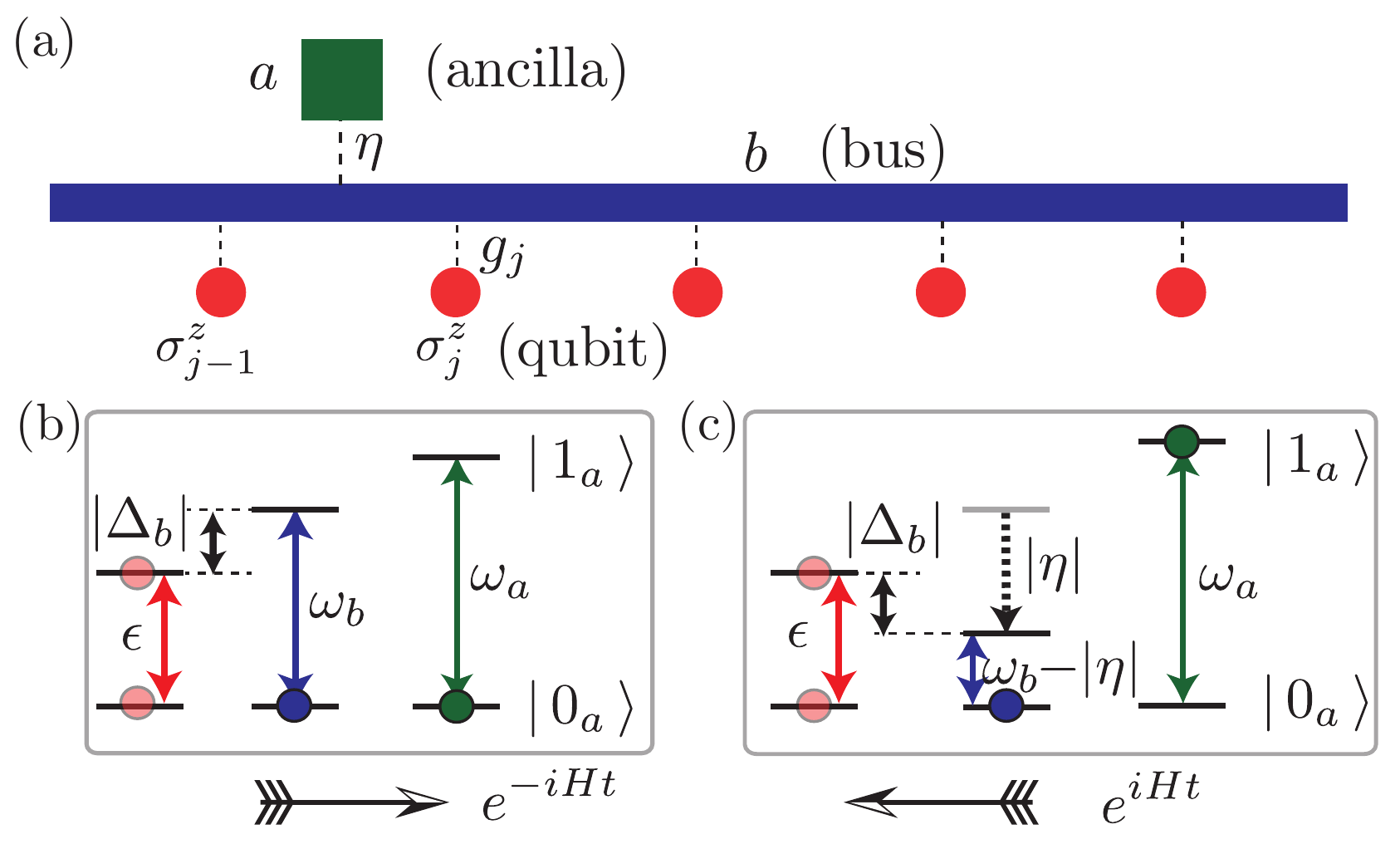}
\caption{Schematic diagrams of a cavity-QED implementation of an all-to-all coupled spin model.  (a) Illustration of the many-body system, consisting of system qubits/spins (red circle), a coupling cavity (blue bar) serving as a passive quantum bus, and an ancilla cavity (green box) serving as a quantum clock.   (b) When there is no photon in the ancilla cavity,  the coupling cavity frequency $\omega_b$ is above the qubit frequency $\epsilon$, with a negative detuning $\Delta_b < 0$.  (c) When there is one photon in the ancilla cavity, the coupling cavity frequency $\omega+\eta$ (where $\eta <0$) is pushed down below the qubit frequency $\epsilon$ by a distance $\abs{\Delta_b}$, which inverts the sign of the detuning and hence the sign of the controlled Hamiltonian.  }
\label{nonlocal}
\end{figure}

The ancilla photon number $n_a$ is a good quantum number, since $[a^\dag a, H_s]=0$. For our use of the ancilla, we restrict $H_s$ in the $n_a=0$ and $n_a=1$ sectors. This can be ensured when introducing nonlinearity into the ancilla cavity by imbedding a superconducting qubit/junction into it (see App.~\ref{circuit-QED} for details) \cite{footnote6}.  We can hence divide the system Hamiltonian into the two ancilla sectors, i.e.,  $H_s$$=$$\sum_{n_a=0,1} H_{s, n_a} \ketbra{n_a}$.   The form of $V$ does not depend on the ancilla photon number, while  $H_0$ can be rewritten as
\begin{align}
\non H_{0}   =&\sum_{n_a=0,1} [ (\omega_b + \chi   n_a ) b^\dag b + \omega_a n_a \\
 &+ \frac{1}{2} \epsilon  \sum_j \sigma^z_{j, j+1}  ]  \ketbra{n_a}.
\end{align}
From the above equation, we can see clearly see that the cavity frequency is controlled by the ancilla photon state.   For convenience, we introduce the ancilla-dependent detuning $\Delta_{b, n_a}$$=$$\Delta_b -\eta   n_a$ for both sectors.

We now treat $V$ perturbatively in the dispersive regime ($g_j \ll \abs{\Delta_{b, n_a}}$) for both ancilla sectors, and integrate out the coupling cavity and finally project to the $n_b=0$ sector.  The resulting effective Hamiltonian \cite{Blais:2007hh, Majer:2007em, Douglas:2015hda, lezTudela:2015gd} up to the second order in perturbation theory \cite{footnote4} is 
\begin{align}\label{nonlocalH}
\nonumber H_\text{eff} =& H_0 + \bigg[ \sum_{j,j'} \frac{g_j g_{j'}}{\Delta_{b, n_a}} \sigma^+_j \sigma^-_{j'} + \sum_{j} \frac{1}{2}\frac{g^2_j }{\Delta_{b, n_a}} \sigma^z_j  \bigg] \ketbra{n_a} \\
                     & +\mathcal{O}\left(\frac{g_j^4}{\Delta_{b, n_a}^3}\right).
\end{align}
The first term at the second order is the so-called `quantum-bus' interaction, i.e., the flip-flop interaction mediated by the virtual photon in the coupling cavity bus \cite{Blais:2007hh, Majer:2007em}.  The second term represents the Lamb shift induced by the cavity bus.  The prefactors of both terms depend on the detuning $\Delta_{b, n_a}$, which is controlled by the ancilla state $\ket{n_a}$.   In order to reverse the sign of these prefactors, we chose the cross-Kerr nonlinearity $\eta$ such that $\Delta_{b, 1} = -\Delta_{b, 0} =-\Delta_b $, which leads to the condition:
\be\label{nonlocalcondition}
\eta= 2(\epsilon - \omega_b) \equiv 2 \Delta_b.
\ee
When enforcing this condition, the effective Hamiltonian in the rotating frame with frequency $\epsilon$ can be written as 
\begin{align}\label{nonlocaleff}
\non \tilde{H}_\text{eff} =&  (1-2 a^\dag a) \bigg[ \sum_{j<j'} \frac{g_j g_{j'}}{\Delta_b} (\sigma^+_j \sigma^-_{j'} +\text{H.c.}) + \sum_{j} \frac{1}{2} \frac{g^2_j }{\Delta_b} \sigma^z_j  \bigg] \\
&+\mathcal{O}\left(\frac{g_j^4}{\Delta_b^3}\right).
\end{align}
Here, the effective Hamiltonian has exactly the form suggested in Eq.~\eqref{Htot}, and the `arrow of time' is controlled by the ancilla photon number $a^\dag a = 0$ or $1$ as desired.  As shown above,  the Hamiltonian controlled by the ancilla is an all-to-all coupled XY model in the presence of external field (corresponding to the Lamb shift term).  One can also easily realize disorder in the coupling strengths.  Additional ZZ-interaction arises in the fourth-order perturbation \cite{dicarlo_demonstration_2009, footnote7} [and the embedding of the ancilla is also realized once Eq.~\eqref{nonlocalcondition} is satisfied]:
\be
V_\text{ZZ} =(1-2 a^\dag a)\sum_{j<j'}  \frac{2 g^2_j g^2_{j'}}{\Delta_b^3} \sigma^z_j \sigma^z_{j'}.
\ee 
The ZZ-interaction strength can be made stronger than this if one uses the transmon qubits \cite{koch_charge-insensitive_2007}, where the third-level of transmon contribute significantly to the ZZ interaction \cite{dicarlo_demonstration_2009}.
Finally, we note that the presence of the Lamb shift is crucial for implementing the controlled operations mentioned in Sec.~\ref{general}, as will be explained in detail in Sec.~\ref{circuit-QEDprotocol}.

\subsection{Local model}\label{sec:local}
Now we  discuss the realization of local lattice models. We present a generic cavity-QED array implementation, which can be realized with, e.g., circuit-QED network and superconducting qubit array in a 3D cavity. The central idea is to use a global cavity as the ancilla, which enables both quantum switching of the `arrow of time' and readout of the OTO correlator. The effective target Hamiltonian $H$ we obtain is a spin-1/2  XY spin model.  We also generalize to an extended Bose-Hubbard model and models with spatial disorders in Sec.\ref{sec:generalization}.

\begin{figure}
\includegraphics[width=1\columnwidth]{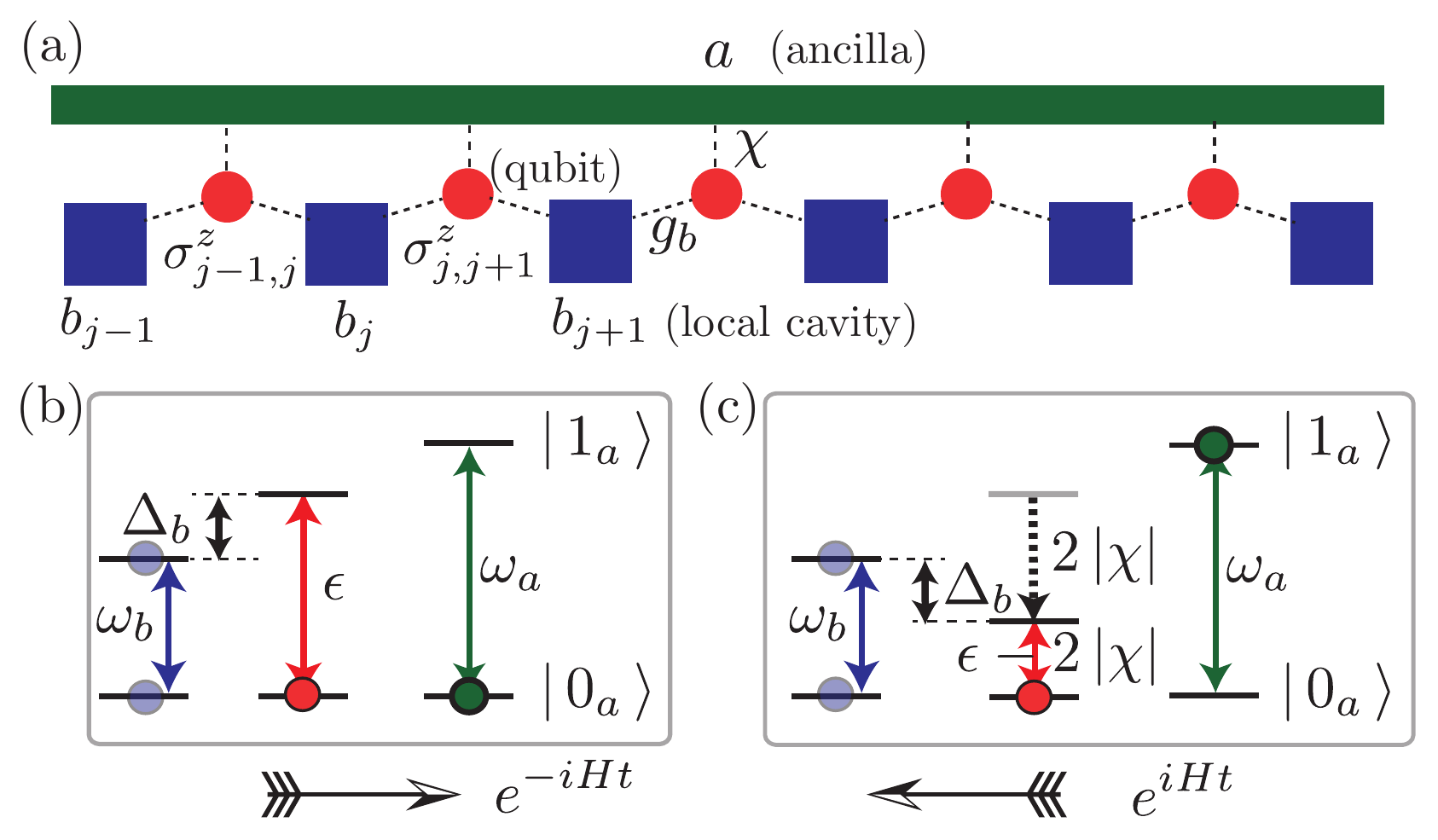}
\caption{Schematic diagrams of the cavity-QED implementation of a local model.  (a) Illustration of the many-body system, consisting of local cavities (blue box), qubits (red circle) mediating interactions between the cavities,  and a global control cavity (green bar) serving as a quantum clock.   (b) When no photon is present in the global control cavity, the qubit energy $\epsilon$ is above the local cavity frequency $\omega_b$, with a positive detuning $\Delta_b$.  (c) When a single photon is present in the global control cavity, the qubit energy $\epsilon' \equiv \epsilon+2\chi$ (where $\chi <0$) is pushed down below the local cavity frequency $\omega_b$ by a distance $\Delta_b$, which inverts the sign of the detuning and hence the sign of the controlled Hamiltonian.  }
\label{cavity-scheme}
\end{figure}

The scheme is illustrated in Fig.~\ref{cavity-scheme}: the blue boxes represent local cavities associated with photon operators $b_j$, which play the role of active degrees of freedom. These local cavities are coupled by intermediate qubits (red circles, associated with Pauli operators $\sigma^z_{j, j+1}$) which are passive degrees of freedom and will be eventually integrated out. Note that this is different than the non-local Hamiltonian in the previous subsection where $\vec{\sigma}$'s were active degrees of freedom while $b_j$ were passive. In addition, similar to the non-local case, the qubits are coupled to a global cavity (described by photon operator $a$), which will serve as the ancilla.   We proceed as before, and split the entire Hamiltonian $H_s$ into two parts, i.e.~$H_s = H_0 + V$:

\begin{align}\label{H0V}
\nonumber H_0 =& \omega_b \sum_j b_j^\dag b_j +  \frac{1}{2} \epsilon \sum_j \sigma^z_{j, j+1} + H_\text{disp},  \\
\nonumber H_\text{disp}=& \chi a^\dag a \sum_j \sigma^z_{j, j+1}, \\
      V =& g_b \sum_j[ b^\dag_j (\sigma^-_{j,j+1}+ \sigma^-_{j-1,j}) + \text{H.c.}].
\end{align}
In place of the cross-Kerr interaction in Eq.~\eqref{eq:Hnonlocal}, $H_0$ now contain a term $H_\text{disp}$ which represents the dispersive interaction between the global cavity ($a$) and the qubits ($\sigma$) with interaction strength $\chi$, and is also sometimes called dispersive shift.    For convenience, we define $\epsilon$ is the renormalized frequency of the qubits, with the Lamb shift due to the global cavity already absorbed into the definition.

We note that the dispersive interaction  $H_\text{disp}$ can arise, e.g., from a Jaynes-Cummings interaction in the dispersive regime \cite{Blais:2007hh, Jiang:2008gs}, where we get the dispersive shift $\chi = g_a^2/\Delta_a$.  Here, $g_a$ is the JC interaction strength and $\Delta_a$ is the detuning between bare qubit ($\sigma$) and global cavity ($a$) frequencies.    For weakly-anharmonic superconducting qubits such as transmons,  the derivation of dispersive interaction can be found in Ref.~\cite{Nigg:2012wh}.

%
 
Similar to the non-local case, the photon number $a^\dag a$ is conserved, and we again restrict to  0- and 1-photon sectors. In the following, we want to eliminate the qubit degrees of freedom ($\sigma$) perturbatively and find an effective Hamiltonian that local cavities ($b$) form an XY model of which the sign is determined by the ancilla photon number.

We consider the dispersive regime where the local cavities and qubits are far detuned in both ancilla sectors, compared to the JC interaction strength, namely 
\[
\Delta_{b, n_a} = \epsilon + n_a \chi -\omega_b \gg g_b \ \ (n_a=0,1).
\]
Here $\Delta_{b, 0} \equiv \Delta_b = \epsilon-\omega$ is the bare detuning in the absence of ancilla photon, while $\Delta_{b, 1} = \Delta_b + 2\chi$ represents the modified detuning in the presence of ancilla photon due to the dispersive shift.  In this regime, since the JC interaction is detuned, there is effectively no exchange of excitations between the local resonators and qubits. This leads to separate conservation of total photon number in the local resonators, $N_b =\sum_j b^\dag_j b_j$, and total qubit excitations $S_z =\sum_j \sigma^z_j$. In particular, we are interested in the low-energy sector that all the qubits have zero excitations, i.e. $\ket{\downarrow\downarrow\downarrow \cdots}$, which corresponds to a projector $P_{S_z=0}$ \cite{footnote8}.  We can adiabatically eliminate the qubits by a Schrieffer-Wolff transformation \cite{schrieffer_relation_1966, Bravyi20112793} in each ancilla sector, namely $H_{\text{eff}, n_a}$$=$$ P_{S_z=0} e^{S_{n_a}} H e^{-S_{n_a}} P_{S_z=0}$.  Such a transformation can be done perturbatively by decomposing the effective Hamiltonian and generator $S$ according to different orders in interactions strength $g_b$, i.e.~ $H_{\text{eff}, n_a} = \sum_m H'^{(m)}_{n_a}$ and $S_{n_a} = \sum_m S^{(m)}_{n_a}$. The first-order generator is given by
 \be
 S^{(1)}_{n_a}  =\frac{g_b}{\Delta_{b, n_a}}  \sum_j b_j (\sigma^+_{j-1,j}+\sigma^+_{j,j+1})- \text{H.c.},
 \ee
which leads to the effective Hamiltonian (up to second order) \cite{footnote4} 
\begin{align}\label{Heff}
\nonumber H_\text{eff} =&   \omega_b \sum_j b_j^\dag b_j  - \sum_{n_a} \frac{g_b^2}{\Delta_{b, n_a}} \sum_{j}  [ (b^\dag_j b_{j+1} + \text{H.c.})   \\
&  + 2 b^\dag_j b_{j}  ] \ketbra{n_a}    +\mathcal{O}\left(\frac{g_b^4}{\Delta_{b, n_a}^3}\right).
\end{align}

We want detunings $\Delta_{b, n_a}$ ($n_a=0, 1$) to have the same magnitude and opposite signs for different occupation number $n_a$.   In other words, we need $\Delta_{b, 1} = -\Delta_{b, 0} = -\Delta_b$.    To achieve this, we simply choose
\be\label{condition}
\chi= -(\epsilon - \omega_b) \equiv - \Delta_b.
\ee
In the situation that the dispersive interaction is realized by Jaynes-Cummings interaction,  i.e. $\chi=g^2_a/\Delta_a$, the above requirement becomes 
\be\label{condition2}
g_a= \sqrt{- \Delta_a \Delta_b},
\ee
meaning the JC interaction interaction strength should be the geometric mean of two detunings with opposite signs. 
The whole scheme is illustrated in Fig.~\ref{cavity-scheme}(b, c), where we have chosen the parameter such that $\omega_b < \epsilon  < \omega_a$, and the dispersive shift $\chi$ is hence negative.

When enforcing the condition in Eq.~\eqref{condition}, the effective Hamiltonian in the rotating frame with frequency $\omega_b$ can be written as 
\begin{align}\label{Hefftilde}
\nonumber \tilde{H}_\text{eff} =&  -  \frac{g_b^2}{\Delta_b} (1 -2 a^\dag a) \sum_{j}  [ (b^\dag_j b_{j+1} + \text{H.c.})    + 2 b^\dag_j b_{j}   ] \\
  & +\mathcal{O}\left(\frac{g_b^4}{\Delta_b^3}\right).
\end{align}
Here, the effective Hamiltonian has exactly the form suggested in Eq.~\eqref{Htot}, and the `arrow of time' is controlled by the ancilla photon number $a^\dag a = 0$ or $1$ as desired.    In addition, one can introduce strong optical nonlinearity to the local cavities, by embedding qubits into it.   In this case, the photons in the cavities can be thought as hard-core bosons due to photon blockade \cite{birnbaum_photon_2005, Hoffman:2011fz}, i.e.~$b_j^2=b^{\dag2}_j=0$, as long as the nonlinearity is much larger than the effective hopping strength between the resonators, $g_b^2/\Delta_b$.   Hence $\tilde{H}_\text{eff}$ actually describes an XY-spin model since the hard-core photon is equivalent to a spin-1/2 degree of freedom.      Besides the flip-flop (XY) interaction, there is also a frequency shift with strength $2g_b^2/\Delta_b$ of the photon, of which the sign flips when the `arrow of time' is reversed.  This frequency shift plays a similar role of effective ``magnetic field" applied to the spins in $z$-direction, due to the mapping $Z_j = 2 b^\dag_j b_j -1$.    The shift is crucial because we need a conditional operation as mentioned in Sec.~\ref{general}, which only act on the system evolving either forward or backward in time.  This will be explained in detail in Sec.~\ref{circuit-QEDprotocol}.  The detailed architecture of a circuit-QED network and superconducting qubit array in a 3D cavity is explained in App.~\ref{circuit-QED}, and the numerical diagonalization of the original [Eq.\eqref{H0V}] and effective [Eq.\eqref{Heff} and \eqref{Hefftilde}] Hamiltonian are compared in App.~\ref{sec:comparison}.

\section{Quantum optical realization}\label{circuit-QEDprotocol}

We briefly present the potential realization of models presented in the previous section and argue that such models could be implemented with current technology. While such models can be realized in most of quantum simulation platforms, ranging from cavity quantum electrodynamics (QED) \cite{birnbaum_photon_2005, Jiang:2008gs, Douglas:2015hda, lezTudela:2015gd} to Rydberg atoms \cite{Lukin:2003ct, Saffman:2010ky, Anonymous:s6xSQwvw, Sommer:2015ur}, and trapped ions \cite{Kim:2010ib} systems, motivated by recent advances in superconducting circuits, we focus our discussion on circuit-QED architecture.

Specifically, we  consider a 2D on-chip circuit-QED quantum simulator consisting of hybrid resonator-qubit network, pioneered by a series of  proposals, experiments \cite{houck2012, koch_time-reversal_2010, Hoffman:2011fz, underwood2012, Schmidt:2013us, Zhu:2013cm, Raftery:2014jk, Chiesa:2015vo}. As schematically shown in Fig.~\ref{circuit}(a), each component of our model could be implemented as follows:  the superconducting qubits (red),  the local superconducting transmission-line resonators (blue cylinders), and the global transmission-line resonator (green).   The details about this architecture and an alternative realization with qubit array in 3D superconducting cavity can be found in App.~\ref{circuit-QED}.

\begin{figure}
\includegraphics[width=0.8\columnwidth]{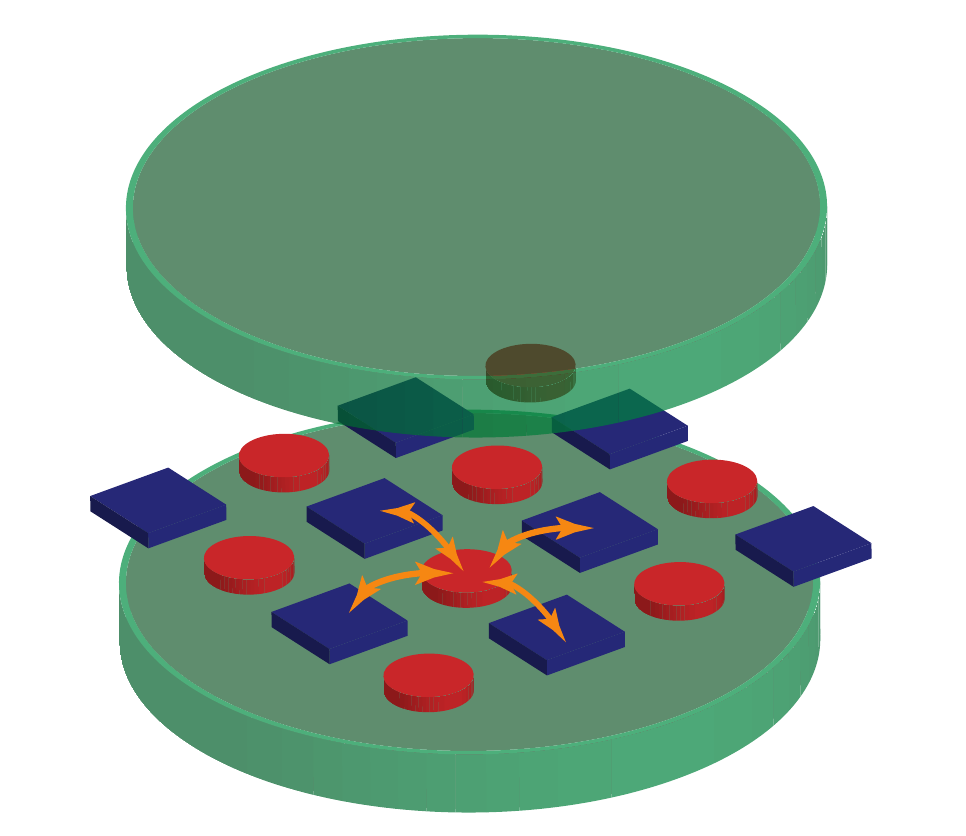}\label{chekcerboard}
\caption{A 2D generalization of the cavity-QED implementation.  Two types of multi-level atoms (qudits), represented by blue boxes and red circles,  form a checkerboard lattice which is placed in a 3D cavity. The blue atoms play the role of active degrees of freedom, while the red atoms are passive coupler mediating interactions between red atoms.  The two types of atoms are coupled by nearest-neighbor flip-flop interactions.  The cavity is selectively coupled to only the red atoms with dispersive interaction to shift their frequencies.   }
\end{figure}

The parameter regime required to implement our model (Sec.~\ref{sec:local}) is within the reach of current technology. The typical qubit and resonator frequencies can span the range 100MHz-15GHz, and the typical coupling strength ranges from 0 to 400MHz. In particular,  the following hierarchy of parameters for the local model can be realized: $g_b \ll \abs{\Delta_b} \sim \chi $ or equivalently $g_b \ll \abs{\Delta_b} < g_a \ll \abs{\Delta_a} $.   In this case, both the condition for sign flipping [Eq.~\eqref{condition} or \eqref{condition2}] and the requirement of dispersive regime can be satisfied.   Similarly, the conditions for implementing the non-local model is also accessible.

Moreover, within this parameter regime, our approximations to obtain the effective Hamiltonian [Eq.\eqref{Heff}] are valid, as we discuss in App.~\ref{complete}. Specifically, the energy spectrum of the full and the effective Hamiltonian are within 0.1\% of each other (for $g_b/\Delta_b =0.1$).

One other key ingredient in implementing our protocol in Sec.~\ref{general} is the conditional operation 
\[
C_{O_1, 0}= O_1 \otimes \ketbra{1_a} +\mathbb{I}_S \otimes \ketbra{1_a}
\]
that only acts on the branch with ``clock" state $\ket{0_a}$ can be realized with the dispersive shifts.  For the local model discussed in Sec.~\ref{sec:local}, the simplest case is to choose $O_1 =X_{j_1} \equiv b^\dag_{j_1} + b_{j_1}$ (in the 0- and 1-photon subspace), meaning that $C_{O_1, 0}$ becomes a CNOT gate.  The dispersive shift of the local resonators $(2 a^\dag a -1)\cdot 2g_b^2/\Delta_b  $ in Eq.~\eqref{Hefftilde} depends on the global control photon (qubit) state, gives the opportunity to realize a CNOT gate by applying a $\pi$-pulse on the local resonator with frequency $\omega_b -2g_b^2/\Delta_b$.  The hard-core photon state of the local resonator is only flipped in the branch with ancilla state $\ket{0_a}$ due to the resonance condition.   Similarly, a conditional operation $C_{O_1, 1}$ which only accesses the branch with ancilla state $\ket{1_a}$ can be applied when sending a $\pi$-pulse with frequency $\omega_b$. One could achieve an arbitrary conditional single-qubit rotation by sending pulses with one of the two corresponding frequencies.  Similar procedure can be applied to the non-local model discussed in Sec.~\ref{sec:nonlocal}, where the Lamb shift in 
Eq.~\eqref{nonlocaleff} contributed to the conditional operations.

\section{Quantum clock versus classical switch:  imperfection and error analysis}\label{error}

In this section, we analyze the stability of protocol against imperfection in the quantum clock and compare it with a previously proposed measurement scheme based on using a classical switch to control the `arrow of time' \cite{Swingle:2016td}.

The main advantage of our protocol is that we do not change the Hamiltonian in situ, and therefore no statistical error corresponding to the fluctuation of the Hamiltonian will be incurred.

\subsection{Classical switch}

In comparison, we analyze a recently proposed protocol of measuring the same correlator \cite{Swingle:2016td} where a continuous classical switch is used to flip the sign of the Hamiltonian, i.e.~ from $H$ to $-H$.   In this type of protocol,  one flips the overall sign of the Hamiltonian by changing the detuning in the cavity-QED system.  

\begin{figure}
\includegraphics[width=1\columnwidth]{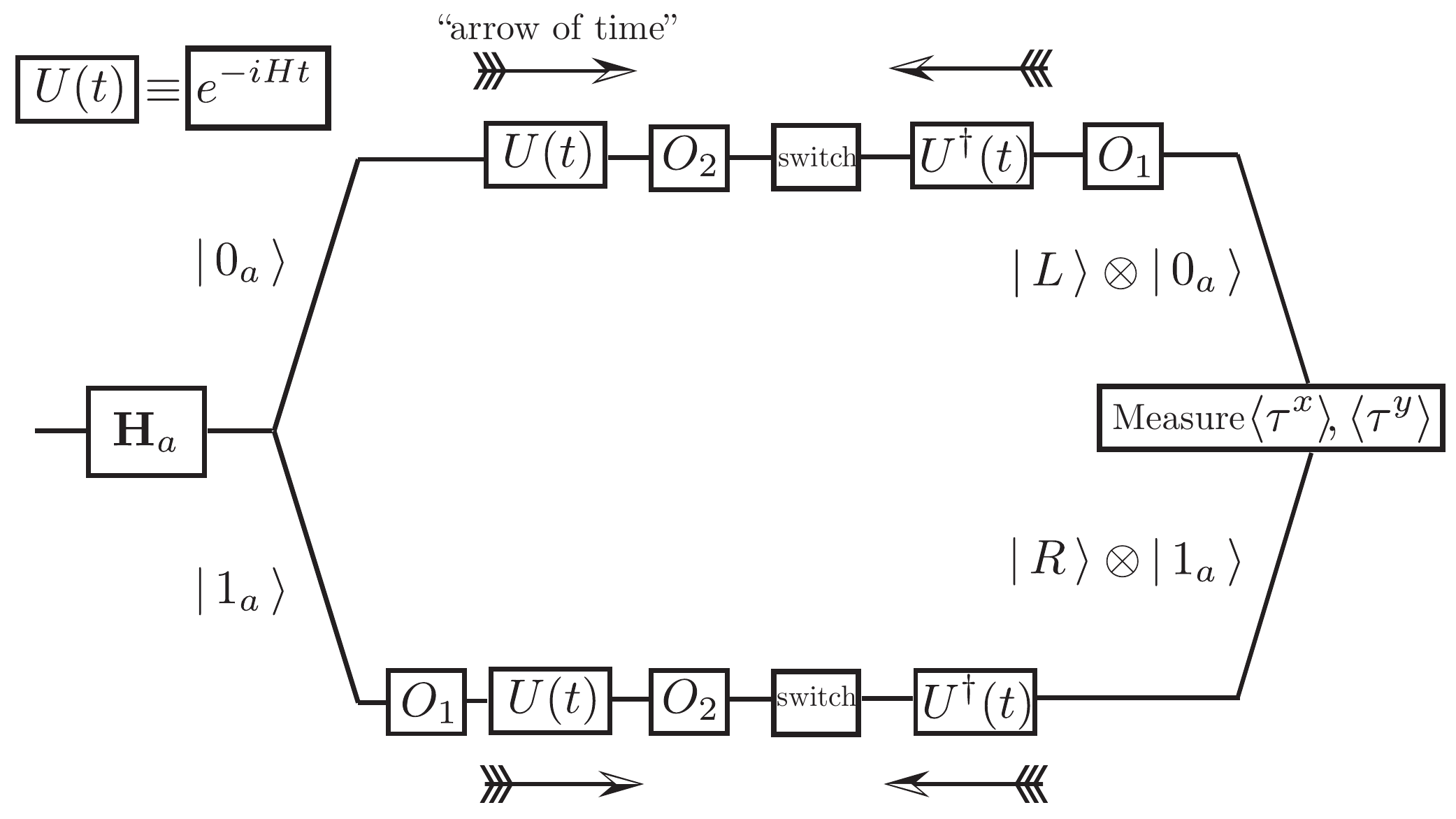}
\caption{Measurement protocol using a classical switch to control the `arrow of time'.  An ancilla qubit is initialized as the superposition of $\ket{0_a}$ and $\ket{1_a}$ and hence split the evolution into two branches in order to do the Ramsey interference.  The ancilla enables conditional-$O_1$ operation but does not control the sign of the Hamiltonian.  Another classical switch (such as the detuning) is used to change the sign of the Hamiltonian and hence flip the `arrow of time'.}
\label{analog}
\end{figure}
\begin{figure}
\includegraphics[width=1.0\columnwidth]{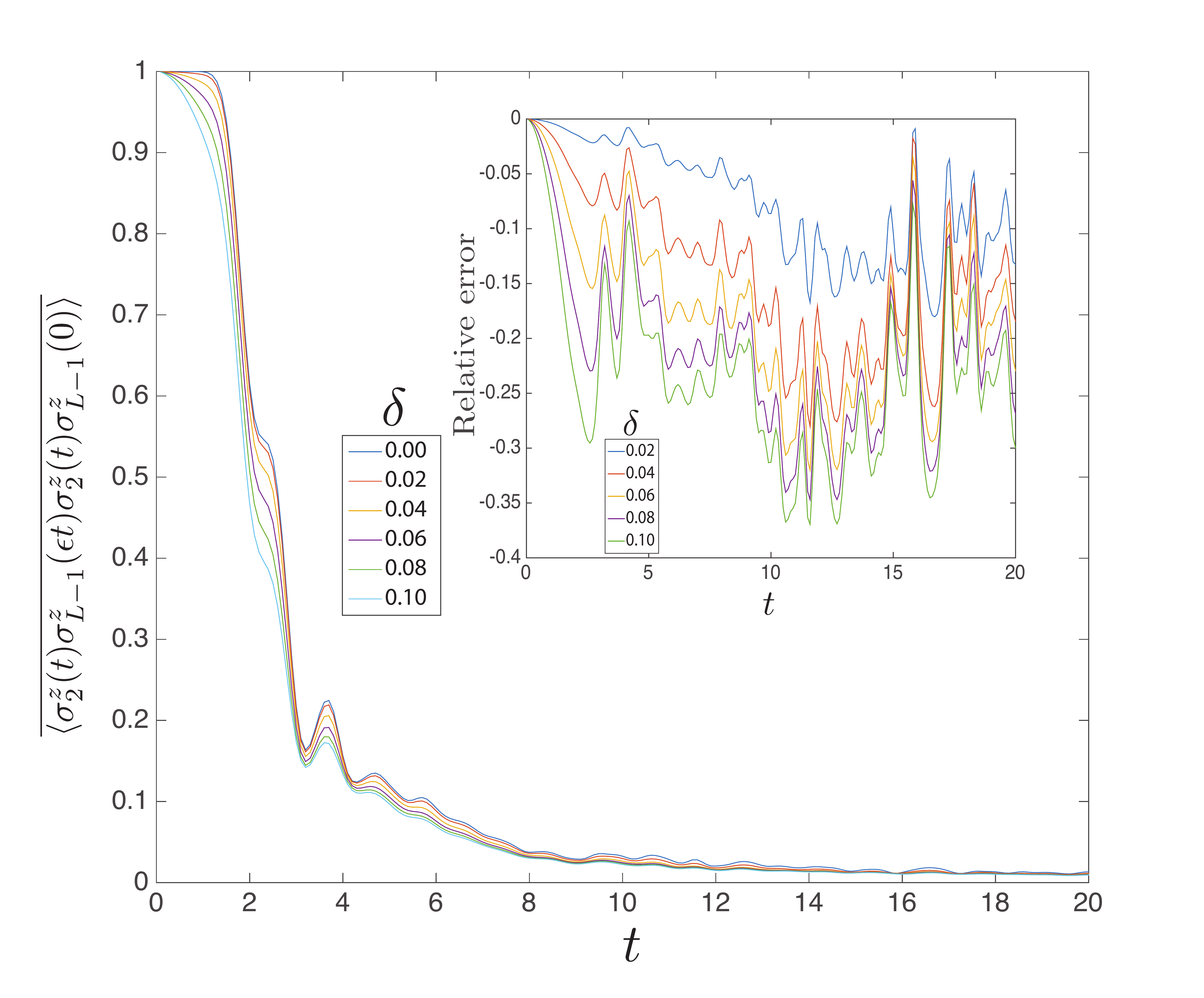}
\caption{Effect of imperfect sign change via classical switch for a spin model. The model considered here is $H$$=$$\sum_i \left( \vec{\sigma}_i.\vec{\sigma}_{i+1} + h_i \sigma^z_i \right)$ where $h_i$ are chosen randomly from a uniform distribution in the interval $[-0.5,0.5]$. The main figure shows the correlator $\langle O_2(t) O_1(t \,\varepsilon)  O_2(t) O_1 (0) \rangle$ with $O_1$$=$$ \sigma^z_2$ and $O_2$$=$$\sigma^z_{L-1}$ where $L$$=$$12$ is the total number of sites. We take $\epsilon$ to be random Gaussian variable with variance $\delta$ and averaging in $\overline{\langle O_2(t) O_1(t \,\varepsilon)  O_2(t) O_1 (0) \rangle}$ is performed over this ensemble. The inset shows the relative error $\left(\langle O_2(t) O_1(t \,\varepsilon)  O_2(t) O_1 (0)\rangle/\langle O_2(t) O_1(0)  O_2(t) O_1 (0)\rangle\right)-1$.} \label{fig:classical_error}
\end{figure}

In order to make the comparison more concrete, we show the protocol with a classical switch from Ref.~\cite{Swingle:2016td} in Fig.~\ref{analog}.   In this protocol, an ancilla qubit is initially prepared in an equal superposition $\frac{1}{\sqrt{2}} ( \ket{0_a} +  \ket{1_a} )$ by the Hadamard gate.   The ancilla is used to perform conditional operation $O_1$, rather than controlling the sign of the Hamiltonian or equivalently the `arrow of time'.   Therefore, the `arrow of time' in both branches always agree with each other.   In the middle of the protocol, the sign of the Hamiltonian is flipped ($H \rightarrow -H$) with a continuous classical switch.  For example, with our setup, one can manually tune the detunings $\Delta_{b, n_a}$ in Eq.~\eqref{Heff} while always staying in the 0-photon sector of the global cavity.   However, since detuning is a continuous variable, the change of the sign cannot be perfect and in fact, the corresponding error can vary from one measurement shot to the next.  Therefore, to estimate the error in OTO, we write the flipped many-body Hamiltonian as  $-(1+\varepsilon) H$, where $\varepsilon$ is a random variable. Correspondingly, the final many-body wavefunctions in the two branches become
\begin{align}
\nonumber \ket{R} =& e^{i H t (1+\varepsilon)} O_2 e^{-i Ht} O_1 \ket{\psi}_S,  \\
  \ket{L} =& O_1 e^{i H t (1+\varepsilon)} O_2 e^{-i Ht} \ket{\psi}_S.
\end{align}
which leads to the following overlap,
\begin{align}
\non & _S\bra{\psi}e^{i Ht} O_2 e^{-i Ht (1+\varepsilon)}  O_1 e^{i Ht (1+\varepsilon)}   O_2 e^{-i Ht} O_1 \ket{\psi}_S \\
\non =& \langle O_2(t) O_1(t \,\varepsilon)  O_2(t) O_1 (0) \rangle
\end{align}
In contrast to the error incurred in the quantum clock, which is independent of the time $t$ (see next subsection), here the error is instead a function of $t \varepsilon$. Thus, the correlation function will match the desired OTO correlator only in the limit $t \varepsilon$ $\ll$ $1$. Fig.\ref{fig:classical_error} shows the error made as a function of $\epsilon$ where we have taken $\epsilon$ to be a random number with mean zero and variance~$\delta$. The error made is clearly time dependent, which potentially makes extracting the functional dependence of the correlator at short times (e.g. to understand scrambling) challenging.

Another potential issue with this approach is that for a local Hamiltonian such as the XY-spin model described above, changing the sign of the total Hamiltonian requires one to change the sign of each individual local term in the Hamiltonian separately, by controlling the detuning \textit{in-situ} site by site, and therefore, it is not obvious how to make the scheme scalable without incurring errors that grow with the system size.   In contrast, a built-in global quantum clock avoids this problem.

\subsection{quantum clock} \label{sec:error_quantum}
Two primary error introduced to our protocol are the imperfection of the pulses (single qubit rotations) acted on the quantum clock (ancilla) and the imperfection in the couplings. The first type of error is generated \textit{in situ}, while the second type is static.  In the following, we analyze the effects of both types of errors.   

\subsubsection{Imperfection in pulses}
Both the initial Hadamard gate ($\pi/2$-pulse) and the $\tau^x$ operation ($\pi$-pulse) which flips the ancilla and hence the `arrow of time' can suffer from errors, since the rotation angles are continuous variables and hence may not be exact.   For a rotation along certain axis $\hat{n}$, we can simply parameterize the rotation error as 
\[
R_{\hat{n}}(\theta + \delta \theta) = e^{-i (\theta + \delta \theta) \hat{n} \cdot \vec{\tau}/2 },
\]  
where $\delta \theta$ is a small random fluctuation which differs in different shots of measurement.   

Assuming the initial Hadamard being perfect, we first consider the imperfection of the two $\tau^x$ flip operations on the ancilla ($\theta_1, \theta_2 =\pi$, $ \hat{n}=\hat{x}$).  Note that due to the two flips of quantum clock divide both upper and lower branches into three sectors, $2^3=8$ paths are generated.  The two paths $\ket{L}$ and $\ket{R}$ are always staying in either of the two branches, i.e. upper $\rightarrow$ upper $\rightarrow$ upper, and lower $\rightarrow$ lower $\rightarrow$ lower respectively, which are the only paths that survive in the absence of error, i.e.~ $\delta \theta_1, \delta \theta_2 =0$.   Once the error is present, the other six paths, which bounce between the upper and lower branches, will have non-zero amplitude.   For example, the path upper $\rightarrow$ lower $\rightarrow$ upper corresponds to the weighted state $(-i \sin \frac{\delta \theta_1}{2})(-i \sin \frac{\delta \theta_2}{2}) [U^\dag (t)]^3 O_2 U^\dag(t) O_1 \ket{\psi}_S$, while the path upper $\rightarrow$ lower $\rightarrow$ lower corresponds to the state $(-i \sin \frac{\delta \theta_1}{2})( \cos \frac{\delta \theta_2}{2}) [U^\dag (t)]^3 U(t)  \ket{\psi}_S$. The errors modify the final state in Eq.~\eqref{final} to
\begin{align}\label{pi_pulse_error}
\nonumber  & \ket{\Psi_f}  =  \frac{1}{\sqrt{2}} \bigg[ \left( \cos \frac{\delta \theta_1 }{2} \cos \frac{\delta \theta_2 }{2} \ket{R} + \sum_{i=1}^3 c_i \ket{E_i} \right) \otimes   \ket{1_a}   \\
 & + \left( \cos \frac{\delta \theta_1 }{2} \cos \frac{\delta \theta_2 }{2} \ket{L} +  \sum_{i=4}^6 c_i \ket{E_i} \right) \otimes   \ket{0_a}  \bigg].
\end{align}
Here, the state $\ket{E_{1,2,3}}$  ($\ket{E_{4,5,6}}$) comes from the other unwanted paths end up in the upper (lower) branch. The amplitudes of them are $c_1=c_4 = -i \sin \frac{\delta \theta_1}{2} \cos \frac{\delta \theta_2}{2}$, $c_2=c_5 = -i \sin \frac{\delta \theta_2}{2} \cos \frac{\delta \theta_1}{2}$  and $c_3=c_6 = \sin \frac{\delta \theta_1}{2}  \sin \frac{\delta \theta_2}{2} $. 

Note that the errors in the $\pi$-pulse do not change the value of the Hamiltonian $H$ and $-H$ for forward and backward propagation.  Nor do the errors change the quantum states $\ket{R}$ and $\ket{L}$, of which the overlap $\bket{L}{R}$ is the OTO correlator.  Now the question is to what extent that our protocol can extract this overlap from the unwanted noise.  When we measure the $\tau^x$ operator according to the protocol, it leads to 
\begin{align}
\nonumber \langle &\tau^x \rangle_f \equiv  \boket{\Psi_f}{\mathbb{I}\otimes\tau^x}{\Psi_f} \\
          =&\cos^2 \left(\frac{\delta \theta_1 }{2}\right) \cos^2 \left( \frac{\delta \theta_2 }{2} \right) \text{Re} [\bket{L}{R}] + \text{Noise}.
\end{align}
The first term is a slightly shrunk signal proportional to the real part of the overlap between $\ket{L}$ and $\ket{R}$.  The second noise term compare from the real or imaginary part of the overlap involving the unwanted paths $\ket{E_i}$.  Since the magnitude of the real or imaginary part of any overlap is bounded by 1, i.e.$\abs{\text{Re} (\text{Im}) \bket{E_i}{E_j} } \le 1, \abs{\text{Re (Im)}  \bket{E_i}{R(L)} } \le 1$, one can derive a bound for the Noise, namely
\begin{align}
\nonumber &\abs{\text{Noise}} \le \abs{\sin \delta \theta_1} + \abs{\sin \delta \theta_2} + \abs{\sin \delta \theta_1}\abs{\sin \delta \theta_2} \\
\nonumber &+ \sin^2 (\frac{\delta \theta_1}{2}) (1+\abs{\sin \delta \theta_2}) + \sin^2 (\frac{\delta \theta_2}{2}) (1+\abs{\sin \delta \theta_1}) \\
\nonumber & +  \sin^2 \frac{\delta \theta_1}{2}  \sin^2 \frac{\delta \theta_2}{2} \\
& = \abs{\sin \delta \theta_1} + \abs{\sin \delta \theta_2} +\mathcal{O}(\delta\theta_1^2 + \delta\theta_2^2+\delta\theta_1 \delta\theta_2).
\end{align}
The above expression suggests that the noise bound is controlled by the errors on the rotation angles. The same prefactor and bound for noise hold for the $\tau^y$ measurement, corresponding to the imaginary part of the overlap.  The signal-to-noise ratio of the overlap has the expression 
\be\label{SNR}
\text{SNR} \approx \frac{ \cos^2 \left(\frac{\delta \theta_1 }{2}\right) \cos^2 \left( \frac{\delta \theta_2 }{2} \right)}{\abs{\sin \delta \theta_1} + \abs{\sin \delta \theta_2}}  \abs{ \bket{L}{R}}, 
\ee
which is also controlled by the error angles and the magnitude of the overlap.   Therefore, the overlap can be resolved once its magnitude is much larger than the noise background.

In addition, the imperfection in the initial Hadamard ($\theta'=\pi/2$, $ \hat{n}=\hat{y}$) leads to the preparation of an unequal superposition of the two branches,
\be
\nonumber \ket{\psi}_S \otimes ( \sqrt{\frac{1-\sin \delta \theta'}{2} } \ket{0_a}+\sqrt{\frac{1 + \sin \delta \theta'}{2} } \ket{1_a}).
\ee
The unequal weight of the wavefunctions in the two branches of the interferometer [conditioned by $\ket{0_a}$ and $\ket{1_a}$ respectively as shown in Eq.~\eqref{pi_pulse_error}] remains in the final output $\ket{\psi}_f$.  
Therefore,  the measurement outcome in the presence of both types errors becomes
\begin{align}
\nonumber \langle \tau^{x(y)} \rangle_f =& \cos \delta \theta' \  \bigg\{\cos^2 \left(\frac{\delta \theta_1 }{2}\right) \cos^2 \left( \frac{\delta \theta_2 }{2} \right) \text{Re(Im)} [\bket{L}{R}] \\
&+\text{Noise}\bigg\}.
\end{align}
An extra prefactor $\cos \delta \theta'$ further shrinks the magnitude of the overlap. On the other hand, the phase of the overlap, i.e.~$\text{Arg}[ \bket{L}{R} ]=\text{arctan} \{  \text{Im} [\bket{L}{R}]/\text{Re} [\bket{L}{R}]\}$ is less affected by the three error angles, since the same prefactors on both the real and imaginary parts cancel with each other.   The SNR ratio remains the same expression as in Eq.~\eqref{SNR} since the same prefactor $\cos \delta \theta'$ is introduced to the noise term.

Last but not least, we emphasize that with the current quantum information technology such as circuit QED, the fidelity of a single-qubit gate can reach over $99.9\%$ \cite{Barends:2014fu}.  Therefore, errors in rotating angles are under control and will not change the order of magnitude of the signal, and we have shown from above that the signal is stable against small imperfection in the gates.

\subsubsection{Imperfection in the couplings}
Before doing the experiments, one needs to tune the parameters such as the detunings $\Delta_a$ and $\Delta_b$ (e.g., by sweeping the magnetic fluxes penetrating the superconducting loops in the SQUID) to satisfy the conditions in Eq.~\eqref{nonlocalcondition}, \eqref{condition} or \eqref{condition2} which allows the reversing of sign exactly.   In addition,  there may be inhomogeneity in the qubit-cavity coupling, namely the coupling strength may have spatial dependence: $g \rightarrow g_{j}, \ g_a \rightarrow g_{a,j}$.  For the nonlocal model discussed in Sec.~\ref{sec:nonlocal} this is not a problem since the inhomogeneity only introduces disorder in the effective coupling strength but not does not affect the condition Eq.~\eqref{nonlocalcondition} which allows exactly flipping the sign with the ancilla. However, for the local model discussed in Sec.~\ref{sec:local},  spatial dependent tunability of the qubit frequency $\epsilon_j$, or equivalently the tunability of detuning $\Delta_{a, j}$ and $\Delta_{b, j}$ is needed to satisfy the required conditions in Eq.~\eqref{condition} or \eqref{condition2}.   Once the tuning is done with high precision, the static imperfection is removed, and no such errors will be introduced \textit{in~situ}.       

The key is to have a calibration procedure that makes sure that the static imperfection is removed or under control. This can be achieved by a simplified version of the Ramsey interference protocol, without applying the operators $O_1$ and $O_2$, such that the cancellation between  the forward and backward evolution could be verified.

\section{Extensions of the local model} \label{sec:generalization}
In Sec.~\ref{sec:local},  we have shown concretely how a 1D XY-spin model can be embedded with a global quantum clock to control the sign of the Hamiltonian.   Here we extend the model in terms of the interaction and lattice type, spatial disorder and dimensionality.

\subsubsection{Soft-core photons and Hubbard model}
Above we focused on hard-core photons which lead to effective spin-1/2 models. Now we consider soft-core photons which allows one to build further interactions. Carrying out  the Schrieffer-Wolff transformation to the 4th-order yields the following correction to the Hamiltonian in Eq.~\eqref{Heff}:

\begin{align}\label{Heff2}
\nonumber \Delta H_\text{eff} =&   \sum_{n_a} \frac{g_b^4}{\Delta^3_{b, n_a}}  \sum_{j} [ 2 b^{\dag}_j b^{\dag}_j b_j b_j  + 6 b^{\dag}_{j} b_{j} b^{\dag}_{j+1} b_{j+1} +8b^\dag_j b_j \\ 
\nonumber &+(2b^\dag_j b_{j+1} +b^\dag_j b_{j+2} + \text{H.c.}) + ({b^\dag}^2_{j+1} b^2_j + \text{H.c.})]   \\
&\cdot \ketbra{n_a}+\mathcal{O}\left(\frac{g_b^6}{\Delta_b^5}\right).
\end{align}
From the above Hamiltonian, we see that all types of interactions, including the newly emerged  on-site interactions, nearest-neighbor density-density interactions, next-nearest neighbor hoppings, and nearest-neighbor pair hoppings all depend on the detuning $\Delta_{b, n_a}$. Therefore, we can easily change the sign of interactions by flipping detuning as we did before, namely using the dispersive shift induced by the global cavity. When imposing the constraint Eq.~\eqref{condition} or \eqref{condition2} as before, the total effective Hamiltonian in the rotating-frame [continuing the series in Eq.~\eqref{Hefftilde}] is 
\begin{align}\label{Hubbard}
\nonumber \tilde{H}_\text{eff} =&     (1 -2 a^\dag a)   \sum_{j} \bigg\{ - \frac{g_b^2}{\Delta_b} [(b^\dag_j b_{j+1} + \text{H.c.}) + 2 b^\dag_j b_j]  \\
\nonumber &+ \frac{g_b^4}{\Delta^3_b}  \sum_{j} [ 2 b^{\dag}_j b^{\dag}_j b_j b_j    +  + 6 b^{\dag}_{j} b_{j} b^{\dag}_{j+1} b_{j+1} +8b^\dag_j b_j \\
\nonumber & +(2b^\dag_j b_{j+1} +b^\dag_j b_{j+2} + \text{H.c.}) + ({b^\dag}^2_{j+1} b^2_j + \text{H.c.})  ]   \\
& +\mathcal{O}\left(\frac{g_b^6}{\Delta_b^5}\right)  \bigg\},
\end{align}
which is actually an extended Bose-Hubbard model with extra pair-hopping terms and an embedded  quantum clock controlling the sign of the Hamiltonian.

\subsubsection{Simulating quenched disorder and localization} \label{sec:disorder}

Above we constructed only the spatially uniform model.   We now note that an XY or extended Hubbard model with spatial disorders in both the hopping strength, and on-site and off-site interactions can also be designed.  
To do so, one simply makes the local JC interaction strength (Eqs.~\eqref{Hefftilde}, \eqref{Hubbard}) spatially disordered, i.e. $g_b \rightarrow g_{b, j}$.  This disorder does not affect the detuning $\Delta_{b, n_a}$ which controls the sign of the Hamiltonian. Hence, the constraint Eqs.~\eqref{condition} or \eqref{condition2} which determine the necessary condition to exactly reverse the sign do not change.

With the spatial disorder in the Hamiltonian, one can potentially realize models with Anderson localization or many-body localization \cite{Oganesyan:2007ex}.  The OTO correlator in these situations may be able to distinguish between a chaotic (ergodic) phase and a many body localized phase.

\subsubsection{Extension in dimensionality and realization}
Generalization of our setup to 2D models is straightforward.    One can devise a checkerboard lattice, with one sub-lattice formed by blue boxes playing the role of active degrees of freedom, and one sub-lattice formed by red circles which will be integrated out and only passively mediate the interactions between blue boxes.   Here one can  go beyond Jaynes-Cummings lattice (oscillator + two level system) and assume that both the blue boxes and red circles represent multi-level atoms (or artificial atoms such as transmons \cite{koch_charge-insensitive_2007}), which can be viewed as qudits, or in simple cases anharmonic oscillators.  The two types of atoms will be detuned from each other and have different level structures, while the interaction between them are of flip-flop (XY) type. The checkerboard lattice is placed in a global 3D-cavity, where the cavity only interacts with the red atoms dispersively and shift their frequencies.  The method of such selective coupling is discussed in App.~\ref{circuit-QED}.    Considering the excitations of active (blue) atoms in the hard-core limit (equivalent to spin-1/2), a similar XY model as Eq.~\eqref{Hefftilde} in 2D can be derived with Schrieffer-Wolff transformation.  This can be easily seen in the limit when the red atom is strongly anharmonic, and therefore can be treated as a two-level system (qubit), thus recovering the results of JC-lattice model. 

Finally we also note that such a checkerboard-lattice setup can also be implemented with Rydberg atoms, where an additional ancilla atom is dispersively coupled only to the sub-lattice serving as passive couplers through the Rydberg blockade mechanism \cite{Saffman:2010ky, Comparat:2010cb, Hofmann:2013gm, Maller:2015is}. Such a partial addressing scheme has been discussed in a recent work about measuring entanglement spectrum with Rydberg atoms \cite{Pichler:2016ua}.

\section{Conclusion and outlook}\label{conc}
In this work, we showed that by embedding a quantum clock into a many-body system, one can control the direction of the time evolution of a many-body system.  One can then use such a quantum clock to measure the out-of-time correlator, which characterizes chaos in a generic quantum many-body system. We have also constructed a class of models implementable in cavity/circuit-QED systems in which such embedding is possible.   In addition, we showed that our protocol which utilizes a quantum clock is robust against imperfection and statistical error in the single-qubit gate, and hence is advantageous over a protocol using a classical switch which is more sensitive to statistical errors.  

Although we focused on realizations with cavity and circuit QED,  the way we construct the models is generic and can be applied to many other platforms where coupling an ancilla qubit globally to the many-body system is possible, such as Rydberg atoms and ion traps.   We also note that the ability to have quantum control of the time evolving direction of a many-body system can have many other  applications, including the ability to measure Loschmidt echo $\equiv {_S\langle} \psi |e^{i H t} e^{-i (H+ \delta H) t}|\psi\rangle_S$, which also requires evolving both backward and forward in time, and is  an alternative measure of quantum chaos. It is also relevant for performing quantum phase estimation, a very useful tool to extract information from a generic quantum simulator without doing quantum-state tomography. From a condensed matter perspective, probing the OTO correlator across the many-body localization-delocalization transition could be very pertinent since the key difference between a thermal phase and a many-body localized phase is precisely that the former is chaotic while the latter is not. As discussed in Sec.\ref{sec:disorder}, this is possible within our setup. Similarly, simulating SYK models \cite{Sachdev:1994, Kitaev:2014} in cavity QED or cold atoms and measuring OTO correlators is another promising direction.

Conceptually, the idea of measuring OTO correlators using the quantum clock is reminiscent of the idea of quantum-controlled ordering of gates discussed in Ref.\cite{Brukner:2014}. The basic result of Ref.\cite{Brukner:2014} is that if in quantum computing, one allows a control switch  that switches the order in which gates are applied (a `permutation switch'), it reduces the computational complexity of certain problems from $O(n^2)$ to $O(n)$. It would be worthwhile to explore the possibility of obtaining such speedups in quantum algorithms using our cavity-QED setup.

\underline{\bf{Acknowledgements:}}

GZ and MH were supported by ONR-YIP, ARO-MURI, AFOSR-MURI, NSF-PFC at the JQI, and the Sloan Foundation.  TG acknowledges startup funds from UCSD and fellowship from the Gordon and Betty Moore Foundation (Grant4304).

\begin{appendix}


\section{The complete sequence of the measurement protocol}\label{sequence}
We recapitulate the steps of our protocol for completeness:

\begin{enumerate}

\item
Initialize the coupled system as $\ket{\psi}_S \otimes \ket{0_a}$.

\item
Apply a Hadamard gate ($\pi/2$-pulse around the $y$-axis) to the ancilla:
\be
\nonumber \mathbb{I}_S \otimes \mathbf{H}_a. 
\ee
\item  Apply a conditional operation 
\be
C_{O_1, 1}= O_1 \otimes \ketbra{1_a} +\mathbb{I}_S \otimes \ketbra{0_a}.
\ee
\item
Conditional evolution for time $t$ 
\be
\non e^{-i H t} \otimes \ketbra{0_a} + e^{i H t} \otimes \ketbra{1_a}
\ee
\item
Apply another conditional operation
\be
C_{O_2, 1}= O_2 \otimes \ketbra{1_a} +\mathbb{I}_S \otimes \ketbra{0_a}.
\ee 
\item
\begin{enumerate}
\item
Apply a $\tau^x$ operator ($\pi$-pulse around the $x$-axis)  to flip the ancilla: 
\be
\nonumber \mathbb{I}_S \otimes \tau^x.
\ee
\item
Conditional evolution for time $2t$ 
\be
\non e^{- 2 i H t} \otimes \ketbra{0_a} + e^{2i H t} \otimes \ketbra{1_a}
\ee
\end{enumerate}
\item
\begin{enumerate}
\item
Apply another $\tau^x$ operator to the ancilla:
\be
\nonumber \mathbb{I}_S \otimes \tau^x.
\ee

\item
Apply another conditional operation:
\be
C_{O_2, 0}= \mathbb{I}_S \otimes \ketbra{1_a} + O_2 \otimes \ketbra{0_a}.
\ee

\item
Conditional evolution for time $t$ 
\be
\non e^{-i H t} \otimes \ketbra{0_a} + e^{i H t} \otimes \ketbra{1_a}.
\ee

\item
Apply another conditional operation:
\be
C_{O_2, 0}= \mathbb{I}_S \otimes \ketbra{1_a} + O_2 \otimes \ketbra{0_a}.
\ee
\end{enumerate}
\item
Measure the expectation value of $\tau^x$ and $\tau^y$ operator under the final state  $\ket{\Psi_f}$ to measure the real and imaginary part of the OTO correlator:
\be
\langle \tau^{x(y)} \rangle_f \equiv \boket{\Psi_f}{\mathbb{I}_S \otimes\tau^{x(y)}}{\Psi_f}  = \text{Re(Im)}[ \bket{L}{R}]. 
\ee
\end{enumerate}

\section{Circuit and cavity QED architecture realizing the local models}\label{circuit-QED}
In this appendix,  we discuss the details about the circuit-QED architecture which realize our desired local model described by Eq.~\eqref{H0V} and illustrated in  Fig.~\ref{cavity-scheme}, and the corresponding experimental protocols.
 
 \subsubsection{2D circuit-QED network}

\begin{figure}
\includegraphics[width=1\columnwidth]{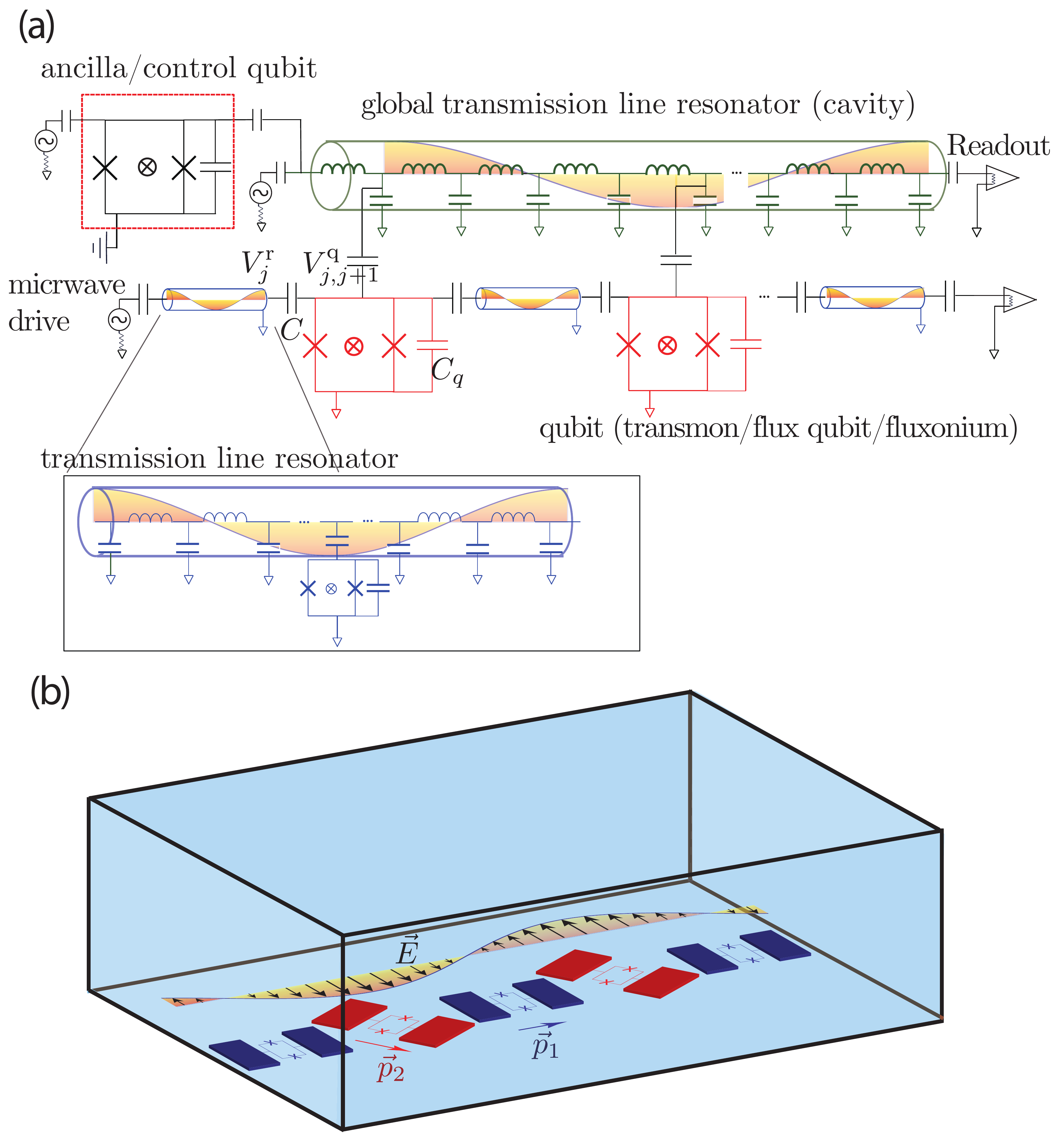}
\caption{Cavity/circuit-QED architecture which realizes the model described by Eq.~\eqref{H0V} and illustrated in  Fig.~\ref{cavity-scheme}. (a) 2D on-chip circuit-QED network.  The setup consists of a global transmission line resonator serving as the `quantum clock', local transmission line resonators which play the role of active degrees of freedom, and qubits which are passive degrees of freedom that mediate interactions between local resonators and are controlled by the global resonator.   Alternatively, one can have an additional ancilla/control qubit coupled to the global resonator, which can either be used to manipulate the cavity photon state, or be dispersively coupled to the local qubits mediated by the cavity bus and hence serves as the `quantum clock'.  (b)  3D cavity-QED with superconducting qubit array, with qubits of two different frequencies (represented as red and blue).  The blue qubits play the role of active degrees of freedom, while the red qubits are passive couplers that mediate interactions between the blue qubits. The dipoles of the qubits are facing different directions to enable selective coupling to the global cavity. }
\label{circuit}
\end{figure}

We first discuss the realization with 2D on-chip circuit-QED network and illustrate it in Fig.~\ref{circuit}(a).  As an example, we show in Fig.~\ref{circuit}(a) the realization of qubits with the Cooper-pair box/transmon, composed of two Josephson junctions and one capacitor.    The level structure and qubit frequency are tuned in situ by the external magnetic flux threading the junction loop. In general, any type of superconducting qubits can be used in the network, such as flux and fluxonium qubits.  

The local $\lambda$-mode transmission-line resonators \cite{footnote5} 
are coupled capacitively to the qubits \cite{houck2012}.  We represent the voltage on the ends of the resonator as $V^\text{r}_{j}$, and the electric charge on the upper superconducting island (non-grounded one) of the qubit as $V^\text{q}_{j, j+1}$.   The capacitive coupling between resonator and qubit on its right leads to the following interaction $T^\text{right}_j= C  V^\text{r}_{j} \cdot  V^\text{q}_{j, j+1}$, where C is the intermediate capacitance.  Canonical quantization allows us to represent the phase variables with creation/annihilation of photon operators, i.e.~$V^\text{r}_{j} = V^\text{r}_{\text{rms}}(b_j+ b^\dag_j)$ and $V^\text{q}_{j, j+1} =e C_\text{g} \cdot \sigma^x_{j, j+1}$, where $V^\text{rms}$ is the root-mean-square Voltage of the resonator, $C_q$ the qubit capacitance, and $e$ the unit charge.  Therefore, with a rotating-wave approximation which drops the counter-rotating term, the interaction can be expressed as the Jaynes-Cummings form $T^\text{right}_j =   g_b  (b^\dag_j \sigma^-_{j,j+1} + \text{H.c.})$, where the JC interaction strength is $g_b = 2e C_q V^\text{r}_{\text{rms}}$. The interaction between the resonator and the qubit on its left has an identical expression.    Sum of all the pairwise interaction terms leads to the realization of the JC interaction $V$ in Eq.~\eqref{H0V}.   Since we eventually need hard-core bosons to simulate spin models, we introduce nonlinearity into the resonators by embedding qubits, which is illustrated in the inset of Fig.~\ref{circuit}(a).  The presence of the qubit inside the resonator leads to photon blockade \cite{Hoffman:2011fz}.

The global transmission-line resonator (cavity) is coupled to all the qubits between the local resonators.  In order to make sure the coupling is uniform, we put the qubits in the peaks (positive or negative) of the resonator mode, implying the length of the resonator is at least $N \lambda/2$, where $\lambda$ is the microwave frequency and $N$ is the total number of qubits.  This also means the control photon occupies the $N^\text{th}$-harmonic mode. Such a super-long transmission-line resonator has been explored experimentally in Ref.~\cite{Sundaresan:2015eu}.  Due to the dressing of the qubit, the level structure of the global resonator also becomes anharmonic, therefore allowing one to manipulate the photon state in the truncated 0- and 1-photon subspace. 

In addition, one could add another ancilla/control qubit coupled to the global cavity.   Instead of exploiting the nonlinearity of the global resonator, one could also use an ancilla qubit to manipulate the photon state through the combination of control-phase gate induced by dispersive interaction and single qubit rotation \cite{Jiang:2008gs}.

An alternative to realize the dispersive-type coupling is to directly couple the ancilla qubit (represented by Pauli operator $\tau$) to all the local qubits, mediated by the virtual photon in the cavity.  In this case, the cavity serves as a quantum bus and hence has no photon occupation. One subtle point is that multiple modes are mediating the dispersive interaction, however, one can select one to play the major role by tuning the qubit frequency close to the frequency of the selected mode.  When the global control qubit is detuned from local qubits, the only interaction survives rotating-wave approximation is the ZZ coupling:
\be\label{ZZ}
 H_\text{disp} \rightarrow H_\text{ZZ}= - \frac{\chi'}{2} \tau^z \sum_j \sigma^z_{j, j+1}.
\ee
Such ZZ interaction is frequently used for a control-phase gate on many platforms.  For example in circuit-QED, such ZZ interaction exists due to the contribution of the third-level of the transmon qubits \cite{dicarlo_demonstration_2009}. One can easily see that by doing the replacement $\tau^z =1- 2 a^\dag a $ in the 0- and 1-photon subspace, the above $H_\text{ZZ}$ is formally identical to $H_\text{disp}$ [in Eq.~\eqref{H0V}] up to a constant frequency shift, which can be absorbed into the renormalized local qubit frequency $\epsilon$.   

An alternative for the global transmission resonator can be a resonator array \cite{koch_time-reversal_2010, houck2012, Schmidt:2013us, Zhu:2013cm}, where we can use the common mode ($k=0$) as the ancilla.  Besides the above approach using capacitive coupling and JC interaction to generate the dispersive interaction perturbatively,  one can also directly couple each resonator in the array to the qubits with a Josephson junction \cite{Jin:2013tz, Nigg:2012wh}.   In this way, the dispersive interaction strength $\chi$ is only proportional to the Josephson energy $E_J$ and does not depend on the detuning in the form of $g_a^2/\Delta_a$, and hence can remain sizable even when the resonator and qubit is far detuned. With this method, the condition Eq.~\eqref{condition} for sign flip is even easier to be satisfied.

 \subsubsection{3D cavity-QED with superconducting qubit array}\label{3d-caivty}

Now we consider a 3D version of the experimental realization.   Instead of considering a hybrid resonator-qubit network as mentioned above, here we only consider a superconducting qubit array in a 3D cavity [c.f. Fig.~\ref{circuit}(b)].  The word `qubit' here is not restricted to two-level systems, but actually refers to multi-level artificial atoms, which is an accurate description for any superconducting qubits, such as transmons \cite{koch_charge-insensitive_2007}.  Experimental realization of a Bose-Hubbard model with transmon array in a 3D cavity has been achieved recently in Ref.~\cite{HacohenGourgy:2015th}.  Still, the array consists of two different types of artificial atoms [illustrated with red and blue in Fig.~\ref{circuit}(b)] with different level structures, achieved for example by choosing different size of the junction loop between the two superconducting islands.   The red qubits play the role of passive couplers that mediate interactions between the blue qubits, consistent with the schematic diagram in Fig.~\ref{cavity-scheme}(a).  

In order to only couple the red qubits but not the blue qubits to the 3D cavity, we exploit the directional property of dipole coupling and so choose different orientations of the red and blue qubits.   As illustrated in Fig.~\ref{circuit}(b), the dipole of the blue qubits $\vec{p}_1$, originating from the Cooper pair tunneling between the two islands,  is perpendicular to the cavity electric field $\vec{E}$.  Therefore, the dipole interaction for the blue qubits $\vec{p}_1 \cdot \vec{E}$ is zero.   On the other hand, the dipole of the red qubits $\vec{p}_2$ is rotated so as not to be perpendicular to the electric field, which in the end gives rise to the dispersive interaction $H_\text{disp}$ in Eq.~\eqref{H0V}.  An alternative trick of realizing selective coupling,  also illustrated in Fig.~\ref{circuit}(b) is by placing the red/blue qubits in the peaks/nodes of the cavity mode.

By treating the two types of qubits (artificial atoms) as anharmonic oscillators, a similar effective Hamiltonian as Eq.~\eqref{Hefftilde} can be derived with Schrieffer-Wolff transformation up to second order.   The scheme can be easily generalized to 2D, such as the checkerboard lattice shown in Fig.~\ref{chekcerboard}.

\begin{figure*}
\includegraphics[width=1.7\columnwidth]{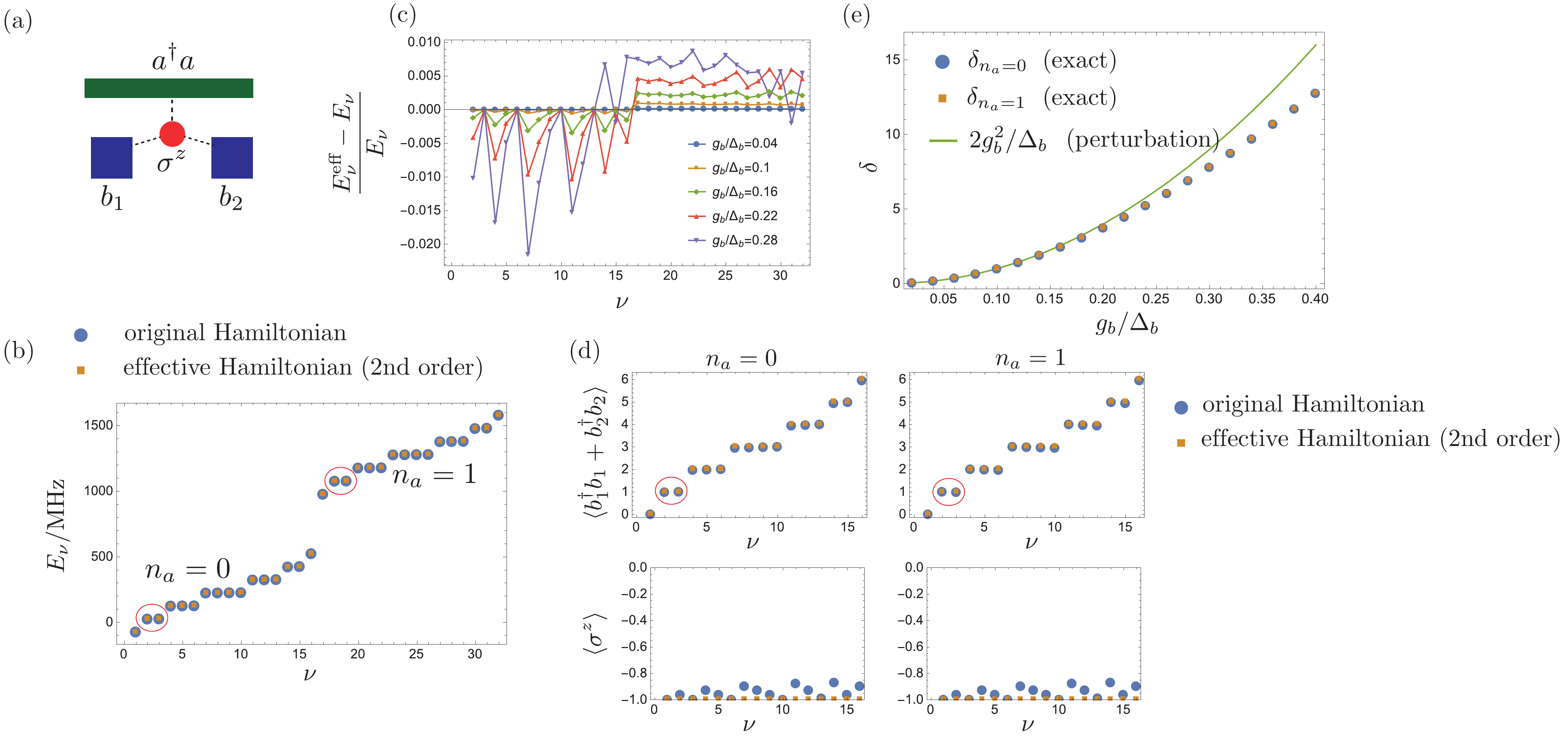}
\caption{Numerical comparison of the original and 2nd-order effective Hamiltonian for a dimer. Parameters: $\Delta_b$$=$$50 \text{MHz}$, $\Delta_a$$=$$-800 \text{MHz}$, and $\chi$$=$$-50\text{MHz}$ (or equivalently $g_a$$=$$200 \text{MHz}$), on-site photon cut-off $n_b^\text{max}=3$. (a) The setup for numerical simulations contains two local cavities, one qubit, and one global cavity. (b) Comparison of the spectrum between the exact (blue circle) and effective (yellow square) Hamiltonian obtained from numerical exact diagonalization.  The spectrum is separate into two ancilla sectors.  The red circle show states in the 1-photon manifold ($\sum_j \langle b^\dag_j b_j \rangle \approx 0, \langle \sigma^z \rangle \approx 0$). (c) The relative error between the exact and effective spectrum for $g_b/\Delta_b=0.1$. (d) The average photon and qubit excitation numbers for the low-lying states in both ancilla sectors, obtained from exact (blue circle) and effective (yellow square) Hamiltonian. The red circles show the states in the 1-photon manifold. (e) The energy splitting in the 1-photon manifold $\delta$ for both ancilla sectors obtained from exact diagonalization of the original Hamiltonian, and the prediction $2g^2_b /\Delta_b$ from second-order perturbation theory.}
\label{two-site}
\end{figure*}

\section{Numerical verification of the effective Hamiltonian}\label{sec:comparison}

In order to verify the effective model we constructed from perturbation theory, we need to compare it from the exact numerical diagonalization of the original Hamiltonian.  In particular, we choose to verify the local model we constructed in Sec.~\ref{sec:local},  which has higher complexity than the all-to-all coupled spin model discussed in Sec.~\ref{sec:nonlocal}.  In this whole section, we compare the numerical diagonalization of the original model Eq.~\eqref{H0V} and the full 2nd-order effective Hamiltonian Eq.~\eqref{Heff} or \eqref{Hefftilde}.

We start with the simplest dimer case as shown in Fig.~\ref{two-site}(a), containing two local cavity sites, and a qubit in between which is coupled to the ancilla cavity. We choose the following specific parameters (which can potentially be realized with circuit-QED systems): $\Delta_b$$=$$50 \text{MHz}$, $\Delta_a$$=$$-800 \text{MHz}$, $\chi$$=$$-50\text{MHz}$ (or equivalently $g_a$$=$$200 \text{MHz}$ if the dispersive interaction arises from the global Jaynes-Cummings interaction perturbatively), and on-site photon cut-off $n_b^\text{max}=3$; we vary $g_b$ in the simulation. In particular, we choose the parameters such that the conditions Eq.~\eqref{condition} and \eqref{condition2} are always satisfied so the sign of the effective Hamiltonian can be flipped by the ancilla.   In panel (b), we compare the spectrum ($E_\nu$) of the original (blue circle) and 2nd-order effective Hamiltonian (yellow squares) at $g_b/\Delta_b =0.1$ ($\Delta_b=50\text{MHz}, g_b=5 \text{MHz}$), which is deep in the dispersive regime and the perturbation is expected to be valid.  The spectrum can be obviously divided into two sectors corresponding to $n_a=0$ and $n_a=1$, and the exact and perturbation results match very well throughout the entire region.    Note that for the original Hamiltonian, we have already selected the spectrum in the subspace with $\langle \sigma^z  \rangle \approx 0$ [see panel (d)] to match the effective Hamiltonian which is restricted in that subspace. The nature of the manifold highlighted by the red circles is to be discussed later in panel (d).  In panel (c), we show the relative error, $(E^\text{eff}_\nu - E_\nu)/E_\nu$, between the exact and perturbation results, with a varying perturbation parameter $g_b/\Delta_b$. Recall that the perturbation is valid in the dispersive regime, with $g_b/\Delta_b \ll 1$. We see the deviation increases with $g_b/\Delta_b$, but still remains small even for sizable $g_b/\Delta_b$, which shows that there is actually a wide parameter region that the perturbation theory is valid.

In panel (d), we plot the average total photon number, $\sum_j \langle b^\dag_j b_j  \rangle$, and qubit excitations, $\langle \sigma^z \rangle$, in both ancilla sectors $n_a=0$ and $1$.  We can see the nature of the pair of states previously circled in panel (c)  are located in the 1-photon ($\langle b^\dag_1 b_1\rangle + \langle b^\dag_2 b_2\rangle \approx 1$) manifold and with zero qubit excitations $\langle \sigma^z \rangle \approx 0$.   We note that the average excitation from the perturbation theory (yellow square) are always exact integers, while the average excitation from the exact results (blue dot) slightly deviates from integer values.  This is due to the fact that, in the effective Hamiltonian of the dispersive regime, total excitations of local cavities ($\sum_j b^\dag_j b_j$) and qubit excitations ($\sigma^z$) conserve separately (i.e., being good quantum numbers). However, this is a consequence of the basis change due to the Schrieffer-Wolff transformation, which effectively rotates the states into dressed basis, where the photon and qubit operators are both dressed operators:  $b_j \rightarrow  e^{S} b_j e^{-S}$ and $\sigma^z \rightarrow  e^{S} \sigma^z e^{-S}$ etc.. Thus, in the original basis, there is still small number of qubit excitation in the sector we label as $\sigma^z=0$ in the dressed basis, vice versa. When $g_b=0$, the circled pairs of states in the 1-photon manifold are doubly degenerate states for both $n_a=0$ and $1$ ancilla sectors, namely $\ket{0_b 1_b}$ and $\ket{1_b 0_b}$.   When $g_b \neq 0$, as predicted by the effective Hamiltonian [Eq.~\eqref{Hefftilde}], there is an effective hopping amplitude $t=(2a^\dag a - 1) g_b^2/\Delta_b$ between neighboring local cavities, which mediated by the intermediate qubit.  The sign of the hopping amplitude changes when the ancilla is flipped, while the magnitude $\abs{t}=g_b^2/\Delta_b$ should remain the same. Therefore, there should be a splitting $\delta =2 \abs{t} = g_b^2/\Delta_b$ between the symmetric and anti-symmetric single-particle states of the dimer, namely $\frac{1}{2}(\ket{0_b 1_b}+\ket{1_b 0_b})$ and $\frac{1}{2}(\ket{0_b 1_b}-\ket{1_b 0_b})$.   In panel (e), we plot the splitting in both ancilla sectors from the exact model, namely $\delta_{n_a=0}$ and $\delta_{n_a=1}$ as a function of $g_b/\Delta_b$,  and compare them with the value $2g_b^2/\Delta_b$ predicted by the perturbation theory.   The match is very good for small $g_b/\Delta_b$ when perturbation theory is valid.   In addition, we note that even when the exact result deviates from the 2nd-order perturbation theory prediction,  the splitting for both ancilla sectors still match.  This fact suggests that our prediction of the equal magnitude of the prefactors in both ancilla sectors may go much beyond the second-order perturbation and may extend to all orders.  In the main text, we already see this to be true for the fourth-order terms in Eq.~\eqref{Heff2} and \eqref{Hubbard}, with the prefactor $(1-2 a^\dag a) g_b^4 /\Delta_b^3 $.  Similarly, for $k^\text{th}$ order perturbation, a prefactor of the form $(1-2 a^\dag a) g_b^k /\Delta_b^{k-1} $ is expected.

\begin{figure*}
\includegraphics[width=2\columnwidth]{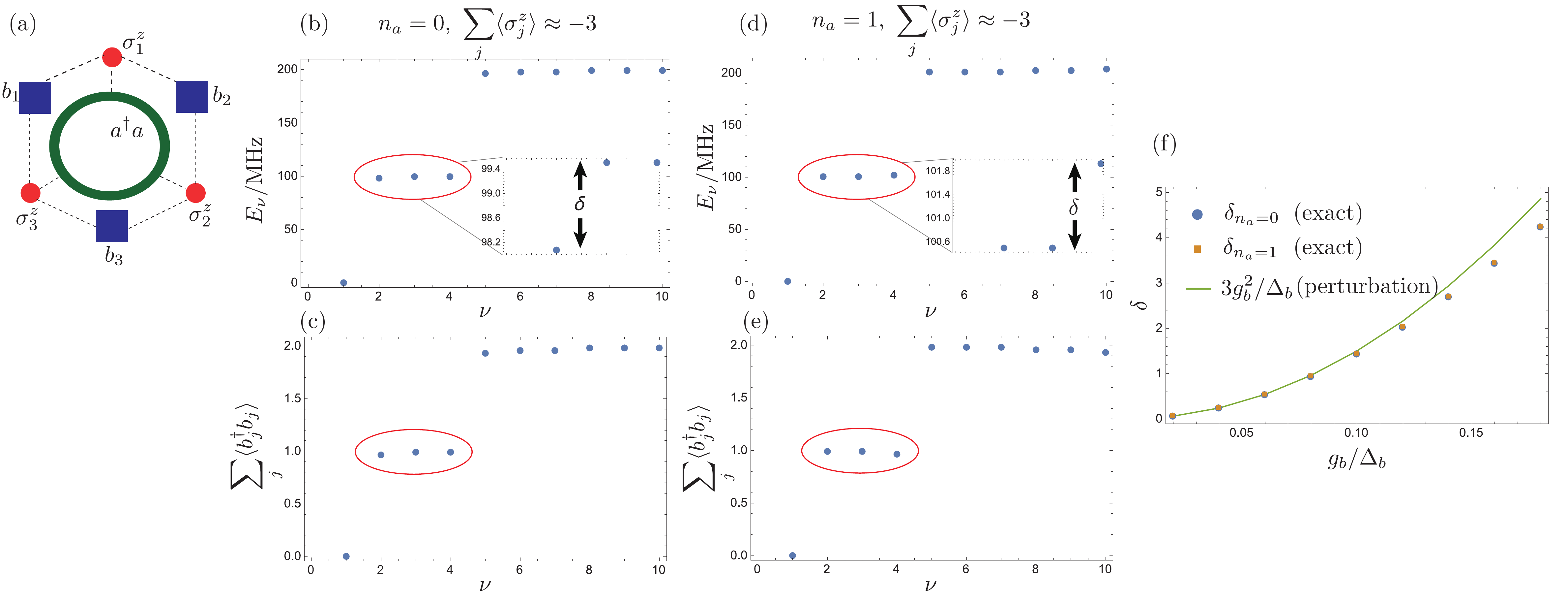}
\caption{Numerical results for a three-site ring. (a) The setup for numerical simulations contains three local cavities, three qubits, and one global cavity, which form a periodic ring. (b-e) The average photon and qubit excitation numbers for the low-lying states in both ancilla sectors, obtained from exact (blue circle) and effective (yellow square) Hamiltonian. The red circles show the states in the 1-photon manifold, and the insets show the zoom-in spectrum in that manifold.  (f) The splitting in the 1-photon manifold for both ancilla sectors obtained from exact diagonalization of the original Hamiltonian, and the prediction $3g^2_b /\Delta_b$ from second-order perturbation theory.}
\label{ring}
\end{figure*}

From the above verification of the dimer case, we see that there is indeed a symmetry of the magnitude of the prefactors in both ancilla sectors.  However, we are not able to check the sign flip induced by the ancilla from the spectrum, since the spectrum of a dimer is invariant under the sign flip of the hopping, which is equivalent to a gauge transformation. 
However, no gauge transformation can flip the hopping signs for a three-site periodic ring, such as the setup shown in Fig.~\ref{ring}(a), which is composed of three local cavities, three qubits in between, and a global ring cavity.   We choose the same parameters and focus still on the 1-photon manifold ($\sum_j \langle b^\dag_j b_j\rangle \approx 1$) of the exact numerical spectrum as shown in panel (b-e).   We can see from the zoom-in insets in panel (b) and (d) that the lowest of the three states in $n_a=0$ sector is singly degenerate, while in the $n_a=1$ sector the lowest states are doubly degenerate.  This can be simply understood by the formula of the effective hopping amplitude $t=(2a^\dag a - 1) g_b^2/\Delta_b$ from Eq.~\eqref{Hefftilde}.  For $n_a=0$ ancilla sector, the effective hopping is $t=-g_b^2/\Delta_b$, which is negative according to the current parameter choice.  In this situation, the spectrum in the 1-photon manifold is $\{-2\abs{t}, \abs{t}, \abs{t}\}$, and the unique ground state in this manifold corresponds to the symmetric state $\frac{1}{\sqrt{3}}(\ket{1_b 0_b 0_b}+\ket{0_b 1_b 0_b}+ \ket{0_b 0_b 1_b})$.   The two degenerate states with higher energy can be chosen as two counter-propagating states with opposite chirality, namely $\frac{1}{\sqrt{3}}(\ket{1_b 0_b 0_b}+ e^{ \pm i 2\pi/3}\ket{0_b 1_b 0_b}+ e^{ \mp i 2\pi/3} \ket{0_b 0_b 1_b})$.  For $n_a=1$ ancilla sector, the effective hopping is $t=g_b^2/\Delta_b$, which is positive and hence leads to frustration of the ring.   In this situation, the spectrum in the 1-photon manifold is $\{-\abs{t}, -\abs{t}, 2\abs{t}\}$, and the doubly-degenerate ground states correspond to the two opposite chiral states, while the symmetric state has higher energy.    Therefore, the signature of sign flipping is clearly shown in the two insets.  In addition, for both ancilla sectors, the splittings ($\delta_{n_a=0}$ and $\delta_{n_a=1}$) between the lower and higher states is fixed to be $3\abs{t} =3 g_b^2/\Delta_b$.   We compare the splittings from the exact diagonalization to the prediction $3 g_b^2/\Delta_b$ from perturbation theory in panel (f) as a function of $g_b/\Delta_b$, and we can see a very good match for small~$g_b/\Delta_b$.    Also, the symmetry of the magnitude of the splitting in both ancilla sectors is again verified.

\section{Complete formula of the second-order effective Hamiltonian}\label{complete}
In the main text, we have derived the effective Hamiltonian of the local model constrained in the sector with zero qubit excitation, i.e.~$\ket{\downarrow\downarrow\downarrow$$\cdots}$, which corresponds to a projection $P_{S_z=0}$.   Here, we release such an constraint, and show the full effective Hamiltonian in the dispersive regime up to second order:
\begin{align}\label{2ndComplete}
\nonumber H_\text{eff} =&  H_0  + \sum_{n_a} \frac{g_b^2}{\Delta_{b, n_a}} \sum_{j}  [ (b^\dag_j b_{j+1} + \text{H.c.})  \sigma^z_{j, j+1}  \\
\nonumber             &+  (\sigma^+_{j-1, j} \sigma^-_{j, j+1} + \text{H.c.})   + \sigma^z_j  \\
              & + b^\dag_j b_{j}  (\sigma^z_{j-1, j} + \sigma^z_{j, j+1} )] \ketbra{n_a}_C 
              +\mathcal{O}\left(\frac{g_b^4}{\Delta_{b, n_a}^3}\right).
\end{align}
We see from the first term that the qubits, like a local quantum switch, mediate qubit-state-dependent hopping of photons on neighboring cavities, which has been previously explored in the context of superconducting circuits \cite{Mariantoni:2008iw}.  On the other hand, the second term shows the flip-flop interaction between neighboring qubits are only mediated by virtual photons (meaning there is no presence of the photon operators), i.e.~the so-called `quantum bus' interaction \cite{Blais:2007hh, Majer:2007em, Zhu:2013cm}.  The third term represents the Lamb shift of the qubits induced by the neighboring local cavities, and the last term represents the dispersive shifts (AC-Stark shifts), which shows the mutual dressing of photons and qubits \cite{Blais:2007hh, Zhu:2013cm}.

\end{appendix}


\begin{thebibliography}{x}
\bibitem{footnote2}
If $O_1, O_2$ were canonically conjugate, then semi-classically
  this correlator is indeed given by $ \left(\frac{d O_2(t)}{d
  O_2(0)}\right)^2$, which is the canonical way to define sensitivity of time
  evolution to the initial conditions.
\bibitem{footnote3} In our convention, the ancilla state $\ket{0_a}$ corresponds to
  the $\ket{\uparrow}$ spin state.
\bibitem{footnote4}  The third-order term is zero in this model. The complete formula
  of the second-order effective Hamiltonian without projecting onto the zero
  qubit excitation subspace is shown in Appendix \ref{complete}.
\bibitem{footnote8} We note that in this situation, if the dispersive interaction
  $H_\text{disp}$ is realized by JC interaction perturbatively, the non-local
  second-order flip-flop interactions between the qubits
  [Eq.~\eqref{nonlocalH}] do not play any role, since all the qubits are in the
  down states. Therefore, the dispersive Hamiltonian $H_\text{disp}$ is indeed
  a valid description.
\bibitem{footnote5} One could also choose the $\lambda /2$ mode, which is the
  fundamental mode of the resonator, leading to an opposite sign of the phase
  variable on the two ends. However, eventually the sign can be gauged out in a
  1D chain.
\end{thebibliography}
\end{document}